\documentclass[useAMS,usenatbib,fleqn]{mn2e}

\usepackage{longtable}
\usepackage{natbib}
\usepackage{graphicx}
\usepackage{times}
\usepackage{amssymb}
\usepackage[greek,english]{babel}
\usepackage{subfigure}
\usepackage{float}

\include{aas_macros.sty}
\newcommand{\galpha}{\greektext a\latintext}
\newcommand{\gbeta}{\greektext b\latintext}
\newcommand{\ggamma}{\greektext g\latintext}

\defcitealias{Bouvier99}{B99}
\defcitealias{Bouvier03}{B03}
\defcitealias{Bouvier07}{B07}
\defcitealias{ASAS}{\emph{ASAS}}

\title[Radiative transfer modelling of AA Tau]{Line and continuum radiative transfer modelling of AA Tau}
\author[Claire F. Esau, Tim J. Harries and Jerome Bouvier]{Claire F. Esau$^{1}$\thanks{E-mail:
claire@astro.ex.ac.uk}, Tim J. Harries$^{1}$ and Jerome
Bouvier$^{2}$\\$^{1}$School of Physics, University of Exeter, Stocker Road, Exeter EX4 4QL\\
$^{2}$UJF-Grenoble 1 / CNRS-INSU, Institut de Plan\'etologie et d'Astrophysique de Grenoble (IPAG) UMR 5274, Grenoble, F-38041, France}
\begin{document}

\date{ }

\pagerange{\pageref{firstpage}--\pageref{lastpage}} \pubyear{2014}

\maketitle

\label{firstpage}

\begin{abstract}
We present photometric and spectroscopic models of the Classical T Tauri star AA Tau. Photometric and spectroscopic variability present in observations of AA Tau is attributed to a magnetically induced warp in the accretion disc, periodically occulting the photosphere on an 8.2 d time-scale. Emission line profiles show signatures of both infall, attributed to magnetospherically accreting material, and outflow. Using the radiative transfer code {\sc torus}, we have investigated the geometry and kinematics of AA Tau's circumstellar disc and outflow, which is modelled here as a disc wind. Photometric models have been used to constrain the aspect ratio of the disc, the offset angle of the magnetosphere dipole with respect to the stellar rotation axis, and the inner radius of the circumstellar disc. Spectroscopic models have been used to constrain the wind and magnetosphere temperatures, wind acceleration parameter, and mass loss rate.
We find that observations are best fitted by models with a mass accretion rate of $5\times10^{-9}$~M$_\odot$~yr$^{-1}$, a dipole offset of between $10^\circ$ and $20^\circ$, a magnetosphere that truncates the disc from 5.2 to 8.8 R$_\star$, a mass-loss-rate to accretion-rate ratio of $\sim0.1$, a magnetosphere temperature of 8500 -- 9000 K, and a disc wind temperature of 8000 K.
\end{abstract}


\section{Introduction}
Classical T Tauri stars (CTTs) are low-mass pre-main-sequence stars. Spectroscopic studies of CTTs show high-velocity redshifted absorption components in their recombination lines, providing evidence of accretion from the circumstellar disc \citep*[e.g.][]{Edwards94,Muzerolle98}. Spectra also show the presence of excess ultraviolet (UV) emission. These can be explained by a magnetospheric accretion model \citep*[][and references therein]{Bertout88,Konigl91,Hartmann94}, in which the magnetosphere of a CTT truncates the circumstellar disc, at which point material attaches to the field lines and falls on to the photosphere. The accreting material travels ballistically and its kinetic energy is liberated as thermal radiation on impact with the photosphere, producing hotspots near the magnetosphere poles. These are the sources of the excess UV emission. Typical accretion rates of CTTs range between 10$^{-7}$ and 10$^{-9}$~M$_\odot$~yr$^{-1}$ (e.g. \citealt{Basri89}).

Spectra of CTTs also show evidence of outflows on different scales, with blueshifted absorption components and blueshifted forbidden line emission present at high velocities \citep{Mundt84,Edwards87}. These signatures are thought to be due to stellar winds, disc winds, and jets. Stellar winds are understood to be powered by a fraction of the energy released at the base of the accretion streams, causing material to escape along open magnetic field lines from the stellar surface \citep[e.g.][]{Matt05}. Disc winds emanate from open field lines threading the accretion disc \citep[e.g.][]{Camenzind90} in a bipolar conical outflow \citep{Konigl00}. This paradigm was first proposed by \citet{Blandford82} to explain observations of jets emanating from the accretion discs of black holes, and was extended by \citet{Pudritz83} to explain bipolar outflows associated with embedded protostars. Highly collimated high-velocity jets, emanating from closer to the star than less collimated disc winds, have also been observed in association with young stellar objects \citep*[e.g.][]{Burrows96,Appenzeller05}. The correlation between mass accretion diagnostics and wind signatures imply that winds are powered by the accretion process \citep{Cabrit90}. Additionally, there is no evidence for mass-loss in the spectra of weak-line T Tauri stars, i.e. when there is no accretion occurring. While stellar winds do appear to contribute to mass-loss, disc winds seem to dominate \citep{Cabrit07}. However, there are numerous possible mechanisms for disc wind formation and the precise origin, or the relative contributions from disc winds of different origins, is still debated (see \citealt*{Ferreira06} for a review).

The magnetospheric accretion paradigm is supported by the results of line models. \citet{Hartmann94} reproduced observed redshifted absorption components and blueshifted emission peaks of Balmer lines using a simple radiative transfer model of magnetospheric infall. This model was extended by \citet{Muzerolle98}, replacing the two-level atom approximation with a multilevel hydrogen atom in statistical equilibrium, followed by a further extension \citep*{Muzerolle01} which included line broadening and sodium line calculations. \citet{Muzerolle01} presented a grid of models across a range of parameter space, varying magnetosphere temperature, line-of-sight inclination, accretion rate, and magnetosphere size. They found H{\galpha} lines which included Stark broadening were more consistent with observations than previous results. They also found instances where line profiles peaked near zero velocity, allowing for natural interpretation of observed CTT spectra that do not show the blueward emission peaks calculated in previous models. While a number of lines were included in the study, H{\gbeta} was the focus of a detailed examination of profile shapes. A similar study focusing on H{\galpha} was carried out by \citet{Kurosawa06} using the radiative transfer code {\sc torus}. This used the same accretion flow model and broadening mechanisms as \citeauthor{Muzerolle01} but the model was extended to include a self-consistent calculation for the hotspot temperature. A disc wind was also included using the formalism of \citet*[][see also \citealt{Long02}]{KWD95} where a biconical wind emerges from a rotating disc. \citeauthor{Kurosawa06} compared results with a classification scheme for H{\galpha} lines proposed by \cite*{Reipurth96}, in which seven classes of line shape were defined depending on the relative strength of the secondary peak to the primary peak (in the case of double-peaked emission lines), and whether the secondary falls blueward or redward of the primary. While some individual profiles were reproducible using models consisting solely of either a disc wind or a magnetospheric accretion flow, \citeauthor{Kurosawa06} found that all classes were readily explained using a hybrid wind-accretion model by varying the angle of inclination to the line of sight, the ratio of mass accretion to loss rates, the wind acceleration rate, and magnetosphere temperature (although one class was better explained using a bipolar outflow or spherical wind, rather than the disc-wind model).

The motivation for this study is to test the observed line formation using an object which has very well constrained physical parameters. AA Tau is an ideal candidate for magnetospheric accretion studies due to its high inclination angle of 75$^\circ$ \citep{Bouvier99}. It is a typical CTT, with a mass of 0.85 M$_\odot$ and a radius of about 1.85 R$_\odot$. Indeed, there have been many observation campaigns involving AA Tau, both photometric and spectroscopic (e.g. \citealt{Bouvier99,Bouvier03}, and the All Sky Automated Survey \emph{ASAS} e.g. \citealt*{ASAS}). Spectropolarimetric observations have also been used to map the magnetic field of AA Tau at different epochs (\citealt{Donati10}). Photometric observations show periodic dips in the light curve of AA Tau, which have been explained by azimuthally asymmetric accretion. If the magnetosphere axis is misaligned with the rotation axis, the two hotspots produced near the magnetosphere poles will sweep in and out of view as the star rotates, separated by half a rotation period. The inner regions of the disc undergo turbulence at the points of interaction with the magnetosphere, causing two waves of material to rise up in disc warps on opposite sides of the star \citep{Bouvier99}. Each warp obscures the photosphere, and since the line-of-sight inclination of AA Tau is sufficiently high, the photosphere and hotspot in the observable hemisphere are only periodically visible. This causes photometric variations, where AA Tau appears fainter during occultation and brighter when the warp is behind the star. AA Tau's photometric period has been shown to vary from 8.2 to 8.6 d \citep*[e.g.][]{Vrba89,Bouvier99,Artemenko12}. This general trend is supported by radial velocity measurements yielding a period of 8.29 d \citep{Bouvier03}, although observations do show more complex variability occasionally, with multiple smaller amplitude dips having been observed by \citet{Bouvier03} and in  \citet[][hereafter \emph{ASAS}]{ASAS}. Conversely, there have also been occurrences where no or very little photometric variability has been apparent over a number of stellar rotation periods. The degree of photometric variability is expected to change with time as the geometry of the system itself will not remain uniform -- the size of the disc warp will vary as the field lines break and reconnect, with short-term variations between rotations due to changes in the local density of material. Photometric variations have been found to reach an amplitude of $\Delta V \sim$ 1.7 mag. Since the photometric period is generally similar to the rotational period of AA Tau, a constraint is placed on the position of the inner edge of the disc, which is equal to the Keplerian corotation radius of 8.8 $R_\star$ for an 8.22 d period. While initially AA Tau's photometric behaviour was regarded as atypical of CTT photometry, light curves showing `AA Tau-like' variations -- that is, variations due to an occulting wall, rather than more periodic variations due to solely the hotspot or completely irregular variations -- have been observed in $\sim$28 per cent of CTTs \citep{Alencar10}, from a sample of 83 CTTs.
AA Tau is a particularly well-constrained object geometrically and shows signatures of azimuthally varying accretion and outflow, both photometrically and spectroscopically, which are consistent with the paradigm of magnetospheric accretion. We are therefore able to use this star-disc-wind system to strongly test the paradigm of magnetospheric accretion in relation to CTTs in general.

Observational photometry from \citet[][hereafter B99, B03, and B07 respectively]{Bouvier99,Bouvier03,Bouvier07} and \citetalias{ASAS} is presented in Section~\ref{sec:photometry} followed by a description of the model calculations and an analysis of the results. Observational spectroscopy from \citetalias{Bouvier07} is presented in Section~\ref{sec:spectroscopy}, followed by a description of the radiative transfer code {\sc torus}. Synthetic H{\galpha}, H{\gbeta} and H{\ggamma} spectra for AA Tau are presented, followed by an analysis of the results. The implications of these results are discussed in Section~\ref{sec:discussion}, followed by a summary in Section~\ref{sec:summary}.

\section{Photometry}
\label{sec:photometry}
The aim here is to use a self-consistent geometrical model of a magnetosphere and an optically thick disc warp to produce synthetic light curves of AA Tau. Models were run using the radiative transfer code {\sc torus} \citep*{Harries00,Symington05}. Various synthetic light curves were produced by varying numerous parameters and the resulting data were compared with photometric observations in order to find the parameter set most consistent with observations. It is well-known that occultation events come and go, with epoch to epoch variations observed across occultations, so we collated as much photometric data as was publicly available to compare with the resulting synthetic light curves. We cannot expect a single parameter set to consistently fit all epochs since the magnetic field of AA Tau varies azimuthally, causing the height of the warp (and hence the depth of the occultation) to vary over time. We are therefore looking for a model which is broadly consistent with the typical deep occultation events observed to occur over an 8.22 d period.

\subsection{Observational photometry}
\begin{figure*}
\centering
{\bf \hspace{10mm} \vspace{2mm} $B$-band, 1995 -- 2005}\\
\includegraphics[width=185mm]{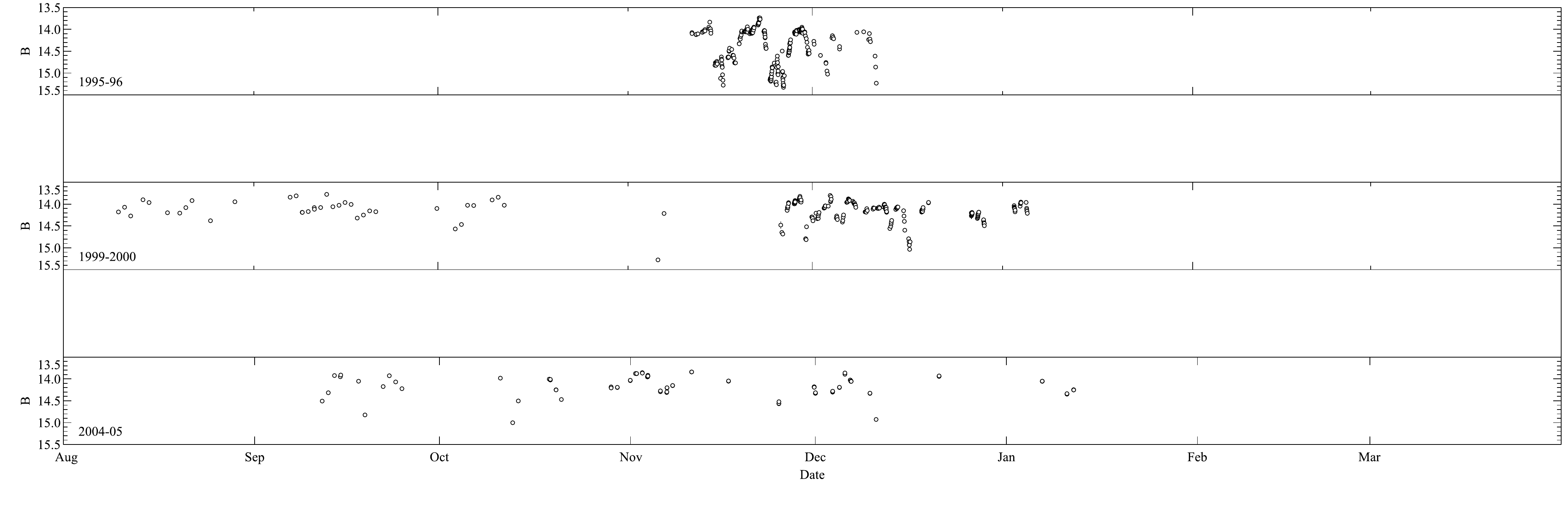}\\
{\bf \hspace{10mm} \vspace{2mm} $V$-band, 1995 -- 2010}\\
\includegraphics[width=185mm]{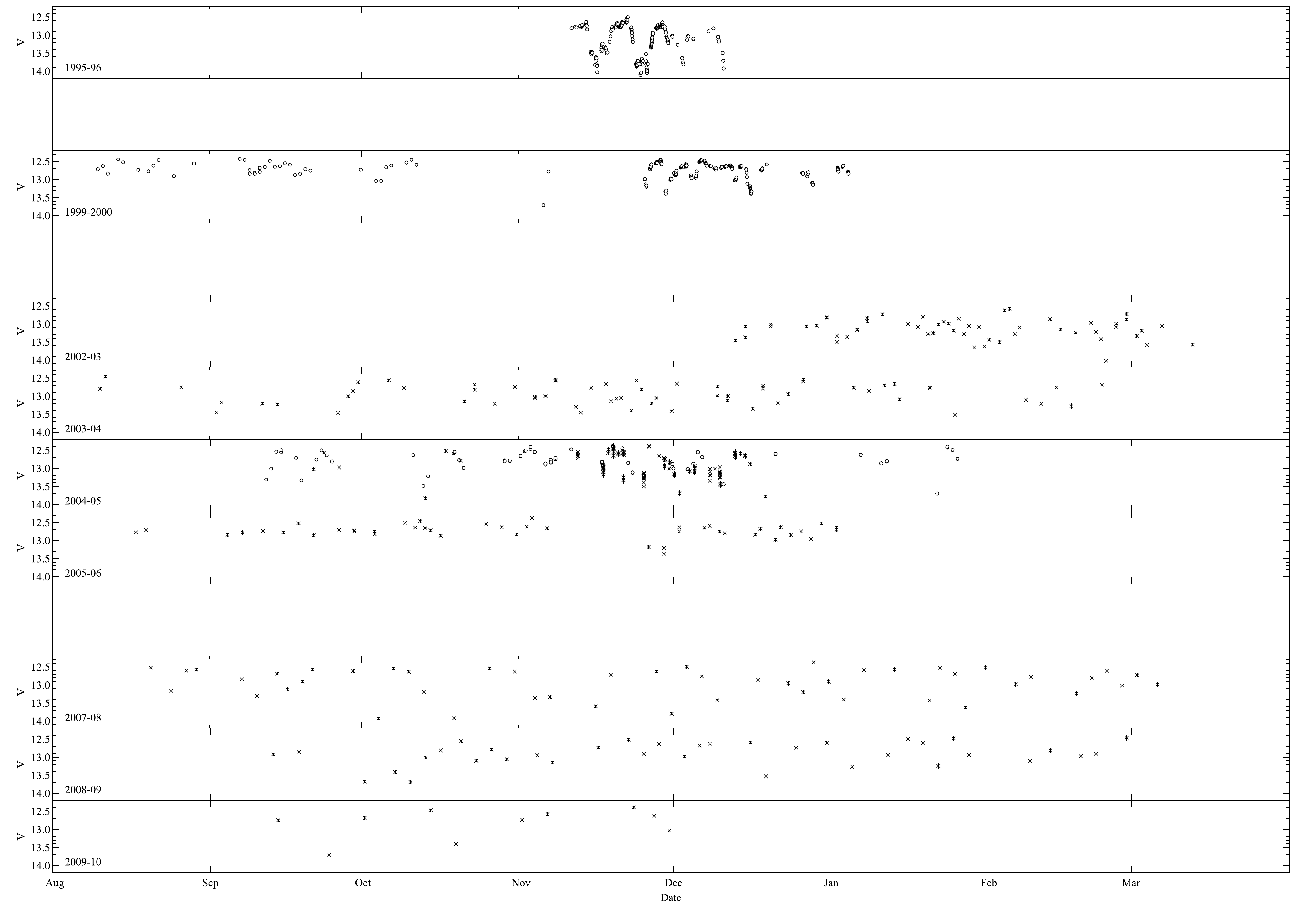}
\caption{All $B$- and $V$-band photometry for AA Tau from 1995 - 2010 \protect\citep{Bouvier99,Bouvier03,Bouvier07,ASAS}. Data from Bouvier et al. is shown with circles and \emph{ASAS} data with crosses. Short-term variations in photometric behaviour are clear, for instance in 1999 -- 2000 where the system shows little variation initially but exhibits deep luminosity dips after just a few weeks. A similar `quiet' period is evident towards the end of 2005.}
\label{allphotometry}
\end{figure*}

$B$- and $V$-band photometric data from \citetalias{Bouvier99} were obtained in November and December 1995, data from \citetalias{Bouvier03} were obtained between August 1999 and January 2000, and data from \citetalias{Bouvier07} were obtained between September 2004 and January 2005. Additional $V$-band data from \citetalias{ASAS} were obtained between December 2002 and November 2009. These have been split into 7 epochs for the photometric analysis, as shown in Table \ref{tab:photobs}. While a number of these epochs do not have sufficient data for a detailed study on variations in the structure of AA Tau over individual rotations, they are still useful in determining the average nature of AA Tau's varying structure over time. All data are plotted in Fig.~\ref{allphotometry}, demonstrating the long-term variation in photometry. One clear source of variation between data sets is the photometric amplitude. \citetalias{Bouvier99} data shows a photometric amplitude of about 1.6 mag in $V$, which decreases to about 1.0 mag in \citetalias{Bouvier03} and \citetalias{Bouvier07} (although \citetalias{Bouvier03} shows two luminosity dips per period, and one instance where a dip disappeared in one cycle). It is obvious from Fig.~\ref{allphotometry} that the data from the first portion of the 1999 observing campaign does not produce well-defined photometric modulation, with variations of about 0.6 mag over a number of weeks. A similar `quiet' spell occurs in 2005 -- 2006, with even less photometric variation. Conversely, data from 2007 -- 2008 show variations returning to about $\Delta V = 1.5$ mag.

\begin{table}
\caption{\label{tab:photobs} \small{Sources of photometric data and the amount of data obtained in each case.}}
\begin{center}
\begin{tabular}{| l | c | c | c |}\hline
Source&Filter&Dates of observations&No. of data points\\ \hline
Bouvier et al. (1999)&&11/11/95 -- 11/12/95&262 ($B$), 275 ($V$)\\
Bouvier et al. (2003)&$BV$&09/08/99 -- 05/01/00&250 ($B$), 273 ($V$)\\
Bouvier et al. (2007)&&11/09/04 -- 25/01/05&99 ($B$), 115 ($V$)\\\\
 &  & 13/12/02 -- 13/03/03 & 55\\ 
 &  & 10/08/03 -- 24/02/04 & 64\\
 &  & 21/09/04 -- 19/12/04 & 101\\
Pojmanski et al.& $V$ & 17/08/05 -- 02/01/06 & 44\\
 (2005) &  & 20/08/07 -- 06/03/08 & 48\\
 &  & 13/09/08 -- 28/02/09 & 36\\
 &  & 14/09/09 -- 27/11/09 & 10\\ \hline
\end{tabular}
\end{center}
\end{table}

\begin{figure}
\centering
\includegraphics[width=85mm]{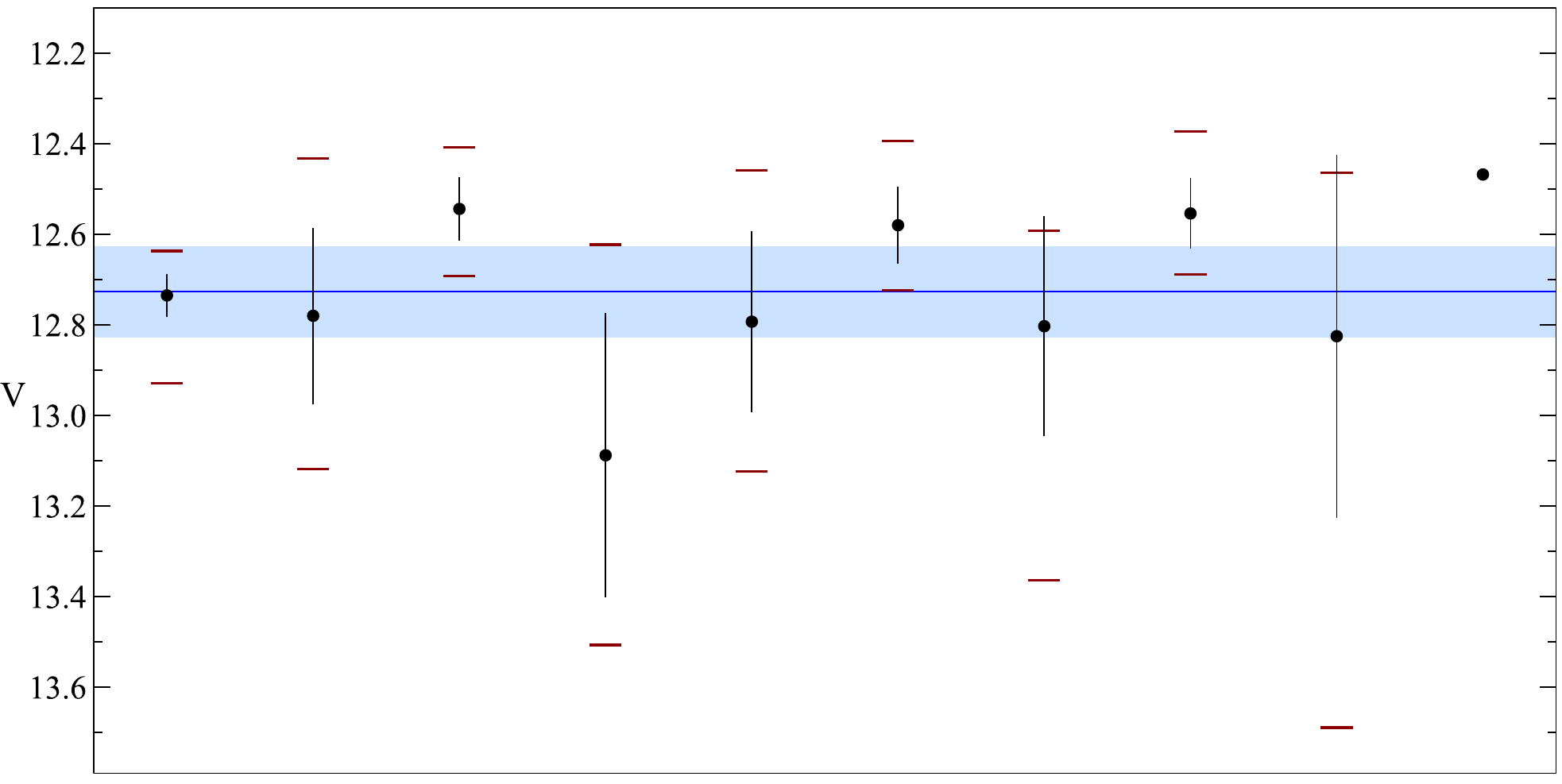}
\caption{Average out-of-occultation magnitudes for each data set. The data points follow the order of \citetalias{Bouvier99}, \citetalias{Bouvier03}, \citetalias{Bouvier07}, and \citetalias{ASAS} 1--7. Note that the 7\textsuperscript{th} \citetalias{ASAS} epoch has no error bars as there was only one data point falling within the defined phases, but has been included for completeness. The global mean is given by the blue line, with the global standard deviation given by the shaded blue region.} 
\label{outofocc}
\end{figure}

In addition to variations in amplitude, AA Tau also exhibits variations in brightness. Fig.~\ref{outofocc} shows the average out-of-occultation $V$-band magnitudes for each data set. Each set of observational data were phased, where phase $\varphi= 0$ is out of occultation and $\varphi = 0.5$ is during occultation. The mean magnitude was calculated for data with $\varphi < 0.1$ and $\varphi > 0.9$ to find the average out-of-occultation magnitude during each epoch, along with the standard deviation of each data set. The global mean and standard deviation are given by the blue line and light blue shaded region, respectively. The global mean lies at $V = 12.73$ while the full spread in data varies from a maximum brightness of $V = 12.37$ (\citetalias{ASAS} 5) to a minimum brightness of $V = 13.70$ (\citetalias{ASAS} 6). The range in individual data sets is denoted by red dashes.

\begin{figure*}
\centering
\includegraphics[width=120mm]{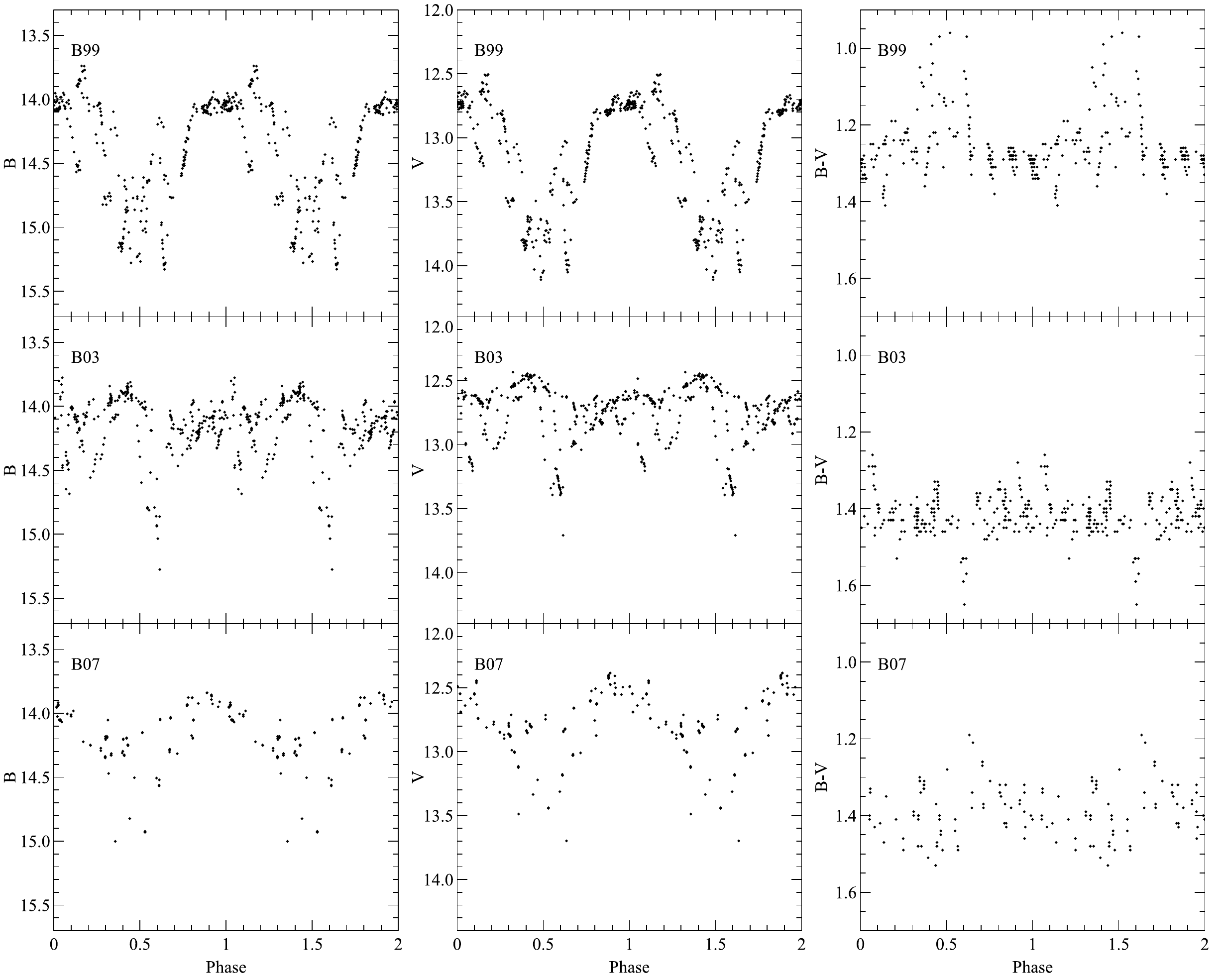}
\caption{$B$ and $V$ photometry for AA Tau from B99, B03 and B07. $B-V$ colours for each data set are on the right. Observations are phased over a period of 8.22 d, where the system is out of occultation at phase $\varphi = 0$ and at maximum occultation at $\varphi = 0.5$. Data from B03 shows two occultation events in one period. This is not an effect of phasing the data as this double trough is also clear in the unphased data.} 
\label{bouvierobs}
\end{figure*}

Colour changes are evident in AA Tau's photometry, with the average ($B$ -- $V$) colour increasing from ${\sim}1.25$ in 1995 to ${\sim}1.42$ in 1999 \citepalias{Bouvier03}. This reddening has been attributed to a lower accretion rate in 1999, causing a reduction in the blue excess. The average ($B$-$V$) colour remained in the region of 1.4 in the 2004 observations, although there is more dispersion. In each case the variations in colour are significantly less than the variations in brightness, with colour variations of about 0.4 mag in 1995 compared to a brightness decrease of about 1.6 mag in $B$ and $V$. Colour variations decreased to about 0.3 mag in 1999 and 2004, with photometric amplitudes of about 1.0 mag in each case. Phased colour plots are presented in Fig.~\ref{bouvierobs}, along with phased $B$ and $V$ observations. While the 1995 colour plot appears show the system being bluer when fainter, there is no evidence of a correlation between colour and brightness in later observations. As pointed out by \citetalias{Bouvier07}, the bluest and reddest colours in the 2004 data both occur during brightness minimum.

Recently AA Tau's photometry has changed significantly, with a decrease in the average brightness level by  ${\sim}2$ mag occurring during 2011 \citep{Bouvier13}.  Observations taken in 2011 show no evidence of coherent photometric modulations, but by the end of 2012 an 8.2 d period is recovered (while remaining faint). $V$-band magnitudes during this period range from $V\simeq$~14 to $V\simeq$~16.5, with an average brightness of $V\simeq$~14.8 and an amplitude of up to 0.9 mag. This has been attributed to a density perturbation in the disc, resulting in an increase in visual extinction from $A_v = 0.8$ mag \citepalias{Bouvier99} to $A_v \ge 4$ mag with no evidence for any significant change in the mass accretion rate.

\subsection{Geometry}
The magnetosphere of AA Tau is modelled as a dipole field along which we assume circumstellar material to be travelling ballistically. This flow is defined by the mass accretion rate $\dot{M}_{acc}$, stellar mass $M_\star$, stellar radius $R_\star$, dipole offset $\theta$, and inner and outer magnetosphere radii, $r_{i}$ and $r_{o}$ respectively. The dipole offset is the angle by which the dipole is tilted with respect to the axis of rotation, where $\theta=0^{\circ}$ describes a system in which the magnetic field and stellar rotation axis are aligned. The system is also inclined to the observer by an angle $i$. The effects that $i$ and $\theta$ have on the observed photometric variations of T Tauri stars have been studied in detail by \citet{MahdaviKenyon}. They also discuss the most likely path that accreting material will take. A tilted dipole results in some field lines providing a longer path to the stellar surface than others. Material in the accretion disc will therefore need to travel further along some lines than others. Material travelling along the longer lines needs to gain potential energy in climbing the line, whereas material travelling from the same position in the disc but over the opposite, shorter field line is able to simply fall on to the star.
Surface hotspots are defined geometrically where the flow hits the stellar surface. We assume that all the kinetic energy from the accretion flow is liberated as thermal radiation. The temperature of these hotspots is calculated from the accretion luminosity and the area of the hotspots. Typically we have found that the hotspots cover about 2 per cent of the stellar surface.

Material is assumed to accrete from around the corotation radius, interior to which the disc is truncated by the magnetic field. $r_{i}$ and $r_{o}$ define the inner and outer radii at which closed magnetic field lines transport material from the disc to the star. In the models presented here, material is assumed to accrete only along energetically favourable field lines, resulting in an azimuthal variation in accretion spot luminosity. The resulting shocks from material hitting the stellar surface produce an accretion signature in the form of a thin elliptical arc near the magnetic pole in each hemisphere.

\begin{figure*}
\centering
\includegraphics[width=160mm]{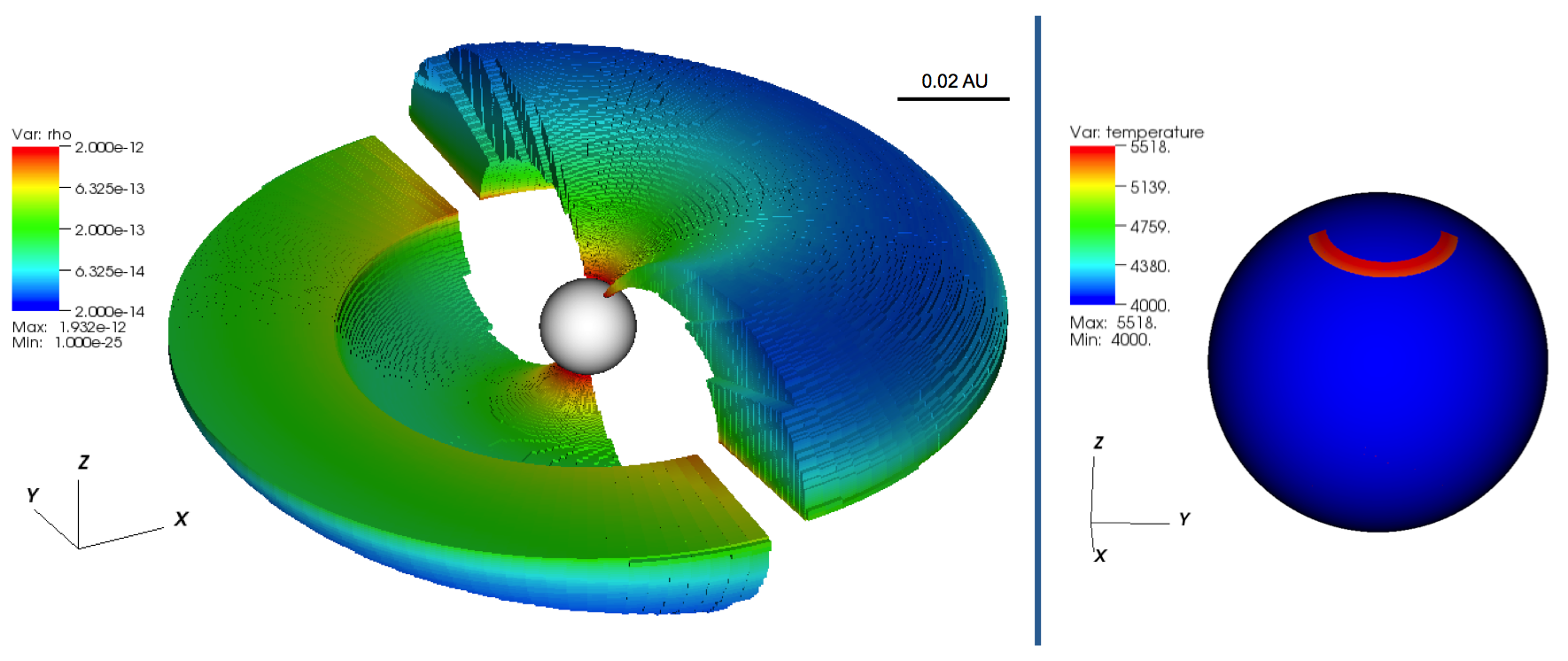}
\caption{AA Tau, modelled with {\sc torus}. The density structure of the magnetosphere is shown on the left with AA Tau shown centrally in white. The density of material covers two orders of magnitude,\,Êincreasing from $2 \times 10^{-14}$ g cm$^{-3}$ (blue) to $2 \times 10^{-12}$ g cm$^{-3}$ (red). The temperature structure of AA Tau is shown on the right, with the hotspot clearly visible.}
\label{aatau}
\end{figure*}

The photometric variations of AA Tau have been attributed to a warp in the circumstellar disc \citepalias{Bouvier99}, caused by the tilt of the magnetic dipole. This produces an optically thick occulting wall. The wall height of the inner disc has been modelled as an azimuthally varying cosine function by \citetalias{Bouvier99},
\begin{equation}\label{eq:wallheight}
h(\phi) = h_\mathrm{max} \left| {\cos\frac{\pi\left(\phi-\phi_{0}\right)}{2}} \right|,
\end{equation}
where $h$ is the height of the disc at azimuth angle $\phi$ and $\phi_0$ is the azimuth at which the disc is at its highest, $h_\mathrm{max}$. This relation has been shown roughly to fit observations. The photometric minima then correspond to times when the highest part of the disc wall occults the star. The occulting warp in our model is defined in a similar manner to \citetalias{Bouvier99}, in that it is parameterized by the aspect ratio $h_\mathrm{max}/r_{o}$. An illustrative model of the disc-star system is shown in Fig.~\ref{aatau}. We do not consider the outer disc structure since this will have negligible effect on the line profiles.

\subsection{Stellar Parameters}
\label{sec:stellparams}
Numerous simulations were run for AA Tau over a range of parameter space. The stellar parameters used in the model are listed in Table \ref{tab:params} with the mass accretion rate ($\dot{M}_{acc}$), dipole offset ($\theta$), aspect ratio ($h_\mathrm{max}/r_{o}$), and inner magnetosphere radius ($r_{i}$), which were varied. Stellar parameters were taken from \citetalias{Bouvier99}, determined as follows: $L_{\star}$ was calculated using the observed $V$-band magnitude of 12.5, when AA Tau is not obscured by the warp, with a bolometric correction of $-0.87\pm0.15$ \citep{KenyonHartmann95}. For a star at a distance of 140~pc, this yields a stellar luminosity of $L_{\star}/L_{\odot}=0.8\pm0.1$. Assuming the standard effective temperature for a K7 dwarf, $T_{eff}=4030\pm30$ K, a stellar radius of $R_{\star}/R_{\odot}=1.85\pm0.15$ was determined. The mass of $M_{\star}=0.85$~M$_{\odot}$ was found from stellar evolution models. A high inclination angle $i$ is apparent, since deep central absorption is present in Balmer emission lines. The source of this absorption is thought to be equatorial, lying at or above one stellar radius from the disc \citep{Kwan97}. The line of sight must intercept both this disc corona and the top of the accretion column in which the Balmer lines are formed in order to explain these observations. A value of $i=75^{\circ}$ has been adopted, yielding a rotation period of 8.22 d using $v$~sin$i=11$~km~s$^{-1}$ \citepalias{Bouvier99}, which is consistent with observations (e.g. \citealt{Vrba89}).

A quadratic limb darkening law from \citet{Kopal50} has been adopted, given by
\begin{equation}\label{eq:limbdark}
I_\mu/I_0 = 1 - a(1 - \mu) - b(1-\mu)^2,
\end{equation}
where $\mu$ is the cosine of the angle subtended by the normal to a given point on the stellar surface and the line of sight, $I_\mu$ is the intensity of the emergent radiation at angle $\cos^{-1}\mu$, and $I_0$ is the intensity of centrally emergent radiation, i.e. at $\mu=1$. $a$ and $b$ are wavelength dependent linear and quadratic limb darkening coefficients respectively. The coefficients used in the model for $B$ and $V$ bands are given in Table \ref{tab:params}, as well as the central wavelengths of these bands.

\begin{table}
\caption{\label{tab:params} \small{Parameters for our grid of models. Stellar parameters $i$, $M_\star$, $R_\star$, $T_{eff}$ and $r_{o}$ are fixed, taken from \citetalias{Bouvier99}. Effective wavelengths and limb darkening coefficients are from \protect\citet{Howarth11}. Model parameters $\dot{M}_{acc}$, $h_\mathrm{max}/r_{o}$, $r_{i}$ and $\theta$ have been fitted using the values listed here.}}
\begin{center}
\begin{tabular}{| l | l | r |}\hline
Parameter & Description & Value\\ \hline
$i$ & Stellar inclination & $75^{\circ}$\\
$M_\star$ & Stellar mass & 0.8 M$_{\odot}$\\
$R_\star$ & Stellar radius & 1.85 R$_{\odot}$\\
$T_{eff}$ & Effective temperature & 4000~K\\
log $g$ & Surface gravity & 3.8\\
$r_{o}$ & Outer magnetosphere radius & 8.8 R$_{\star}$\\
\\
$\lambda_\mathrm{eff}$($B$) & Effective wavelength ($B$ passband) & 4576.4 \AA\\
$\lambda_\mathrm{eff}$($V$) & Effective wavelength ($V$ passband) & 5608.3 \AA\\ 
$a_B$ & Limb darkening coefficient $a$ ($B$-band) & 1.054 \\ 
$b_B$ & Limb darkening coefficient $b$ ($B$-band) & -- 0.165 \\
$a_V$ & Limb darkening coefficient $a$ ($V$-band) & 0.830 \\ 
$b_V$ & Limb darkening coefficient $b$ ($V$-band) & 0.016 \\ 
\\
$\dot{M}_{acc}$ & Accretion rate & $1\times10^{-9}$ \\
&(M$_{\odot}$ yr$^{-1}$)& $2\times10^{-9}$ \\
&&$5\times10^{-9}$ \\
&&$1\times10^{-8}$ \\
$h_\mathrm{max}/r_{o}$ & Aspect ratio & 0.29\\
&& 0.30\\
&&0.31\\
&&0.32\\
&&0.33\\
&&0.34\\
$r_{i}$ & Inner magnetosphere radius & 5.2\\
&(R$\star$)& 6.4\\
&&  7.6\\
$\theta$ & Dipole offset & 10$^{\circ}$\\
&& 20$^{\circ}$\\
&& 30$^{\circ}$\\
&& 40$^{\circ}$\\\hline
\end{tabular}
\end{center}
\end{table}

\subsection{Numerical method}
\label{sec:modelCalcs}

The radiative transfer code {\sc torus} \citep{Harries00,Symington05,Kurosawa06} uses the adaptive mesh refinement (AMR) technique to sample each region of the system being simulated (e.g. magnetosphere, accretion disc) with sufficient resolution, without being too computationally expensive. For 2D models the simulation space is 2D cylindrical and rotationally symmetric in azimuth. Splitting occurs across the cylinder radius $r$ and height $z$ to produce a grid of squares, where each square is the cross-section of a given annulus. 3D simulations use a 3D cylindrical coordinate system in which the top level of the AMR mesh is a cylinder that has a height equal to its radius. Child cells may be formed by either splitting the parent equally in both height, radius, and azimuthal extent (resulting in eight children),  or by just splitting equally  in height and radius (resulting in four children). In general the cells are truncated, cylindrical shells. The principle advantage of this method is that the grid need only be refined in regions where there are significant departures from rotational symmetry, thus minimizing the global number of grid cells.
There are 3 components considered in the photometric models, namely the photosphere, the magnetosphere, and the accretion disc. A disc wind is also included in the spectroscopy models. The setup of these components is described in detail in \citet{Kurosawa06} \cite[see also][]{Symington05}. In summary:

\begin{itemize}
\item[--] The photospheric contribution to the continuum flux is as described in Section \ref{sec:stellparams}. There is an additional contribution from the hotspots that form at the base of the accretion streams due to the thermalized kinetic energy of the infalling gas, which is assumed to radiate away as blackbody emission \citep{Hartmann94}.
\item[--] We use the accretion flow model of \citet[see also \citealt{Muzerolle01}]{Hartmann94}. The density and velocity of the gas at a given point along the field lines are calculated by applying the conservation of mass using the method of \citet{MahdaviKenyon}.
\item[--] The disc wind configuration of \citet{KWD95} is used. This is a biconical geometry parameterized by the mass-loss rate, the degree of collimation, the velocity gradient, and the wind temperature. The azimuthal velocity component  $v_{\phi}$ is calculated simply from the Keplerian rotational velocity at $w_{i}$, where $v_{\phi}(w_{i},z$~=~$0)=(GM_{\star}/w_{i})^{1/2}$, where $w_i$ is the distance between the cylindrical axis $z$ and the emerging point on the disc. Assuming conservation of specific angular momentum, the azimuthal velocity component is $v_{\phi}(w,z) = v_{\phi}(w_{i},0)(w_{i}/w)$, where $w$ is the cylindrical radius. The radial velocity component $v_r$ is calculated using a modified form of the $\beta$ velocity law for hot stellar winds \citep*{CAK75}. The formulation of \citet{CastorLamers79} is used, where the velocity of material at a distance $r$ from the rotational axis $z$ is defined as
\begin{equation}\label{eq:radialvelocity}
v_{r}(r) = v_{0} + (v_{\infty} - v_{0})\left(1-\frac{R_{\star}}{r}\right)^{\beta},
\end{equation}
\noindent where $v_{0}$ is the velocity of material at the base of the wind stream and $v_{\infty}$ is the terminal velocity. $\beta$ is an acceleration parameter where smaller values of $\beta$ yield greater acceleration. 
\end{itemize}

Photometric models are calculated by constructing a grid of rays and integrating a formal solution to the equation of radiative transfer along multiple lines of sight. The rays are chosen to sample the photosphere, the hotspots, and the disc wall with sufficient coverage for each component of the system.
For the spectroscopic models, once the grid has been constructed the solution to the equation of statistical equilibrium from \citet{KleinCastor78} is computed for pure hydrogen under the Sobolev approximation \citep[see also][]{Rybicki78,Hartmann94}. We solve the rate equations using the method detailed in \citet{Symington05} and \citet{Kurosawa06}. Line profiles are then computed by a ray tracing technique, with observed flux calculated by performing a formal integral of the specific intensity at the outer boundary of the simulation space, in the direction of the observer, over all frequencies. 

\subsection{Synthetic photometry}
Monochromatic flux $F$ calculated in the model was converted to apparent magnitude $m$ using 
\begin{equation}\label{eq:apmag}
m = -2.5\,\mathrm{log}\left(\frac{F}{F_0}\right)
\end{equation}
\noindent where $F_0$ is a normalizing flux, corresponding to $m = 0$, with $F_0=6.4\times10^{-9}$~erg~s$^{-1}$~cm$^{-2}$~\AA$^{-1}$ for the $B$-band and $F_0=3.75\times10^{-9}$~erg~s$^{-1}$~cm$^{-2}$~\AA$^{-1}$ for the $V$-band \citep{Cox00}.
The resulting photometry was folded over an 8.22 d period  \citep{Bouvier07} and reddened by $A_V=$0.78 \citepalias{Bouvier99} or $A_B=$1.03, adopting a reddening constant of 3.1 \citep{SchultzWiemer}, for direct comparison with observations. The point of maximum occultation has been set to occur at a phase of 0.5, with no occultation at phase 0.
Ideally a statistical analysis using a $\chi^2$ minimalization technique would have been performed to determine the best fitting parameters. However, due to the variability of AA Tau's photometry, we have taken a pragmatic approach in selecting a best fitting model, judging the level of agreement between models and observations over different epochs by eye.

\citetalias{Bouvier99} calculated an accretion luminosity of $L_{\rm spots}=6.5\times10^{-2}$~L$_\odot$ using the blue excess determined from photometric observations, where
\begin{equation}\label{eq:Lspot}
L_{\rm spots} = A_{\rm spots}\sigma T_{\rm spot}^{4}.
\end{equation}
$A_{\rm spots}$ is the area of the stellar surface which is covered by the accretion spots, and $T_{\rm spot}$ is the spot temperature, found using the $B$-band flux of a spot measured from AA Tau's spectral energy distribution. The \citetalias{Bouvier99} accretion luminosity was used to determine the best mass accretion rate from the models presented here. The range of accretion luminosities calculated in different models for each accretion rate is given in Table \ref{tab:Lacc}. An accretion rate of $\dot{M}_{acc} = 5 \times 10^{-9}$ {M}$_{\odot}$ {yr}$^{-1}$ has been selected, being the accretion rate most consistent with the accretion luminosity from \citetalias{Bouvier99}.

\begin{table}
\caption{\label{tab:Lacc} \small{The range of accretion luminosities calculated for different mass accretion rates}}
\begin{center}
\begin{tabular}{ c | c | c |}\hline
$\dot{M}_{acc} \,(\mathrm{M}_{\odot} \, \mathrm{yr}^{-1})$ & $L_{min} \, (10^{-2}\, \mathrm{L}_\odot)$ & $L_\mathrm{max} \, (10^{-2} \, \mathrm{L}_\odot)$\\
\hline
$1\times10^{-8}$ & 10.9 & 12.5\\
$5\times10^{-9}$ & 5.5 & 6.1\\
$2\times10^{-9}$ & 2.1 & 2.5\\
$1\times10^{-9}$ & 1.1 & 1.2\\
\hline
\end{tabular}
\end{center}
\end{table}

\subsection{Results}

\begin{figure}
\centering
\includegraphics[width=85mm]{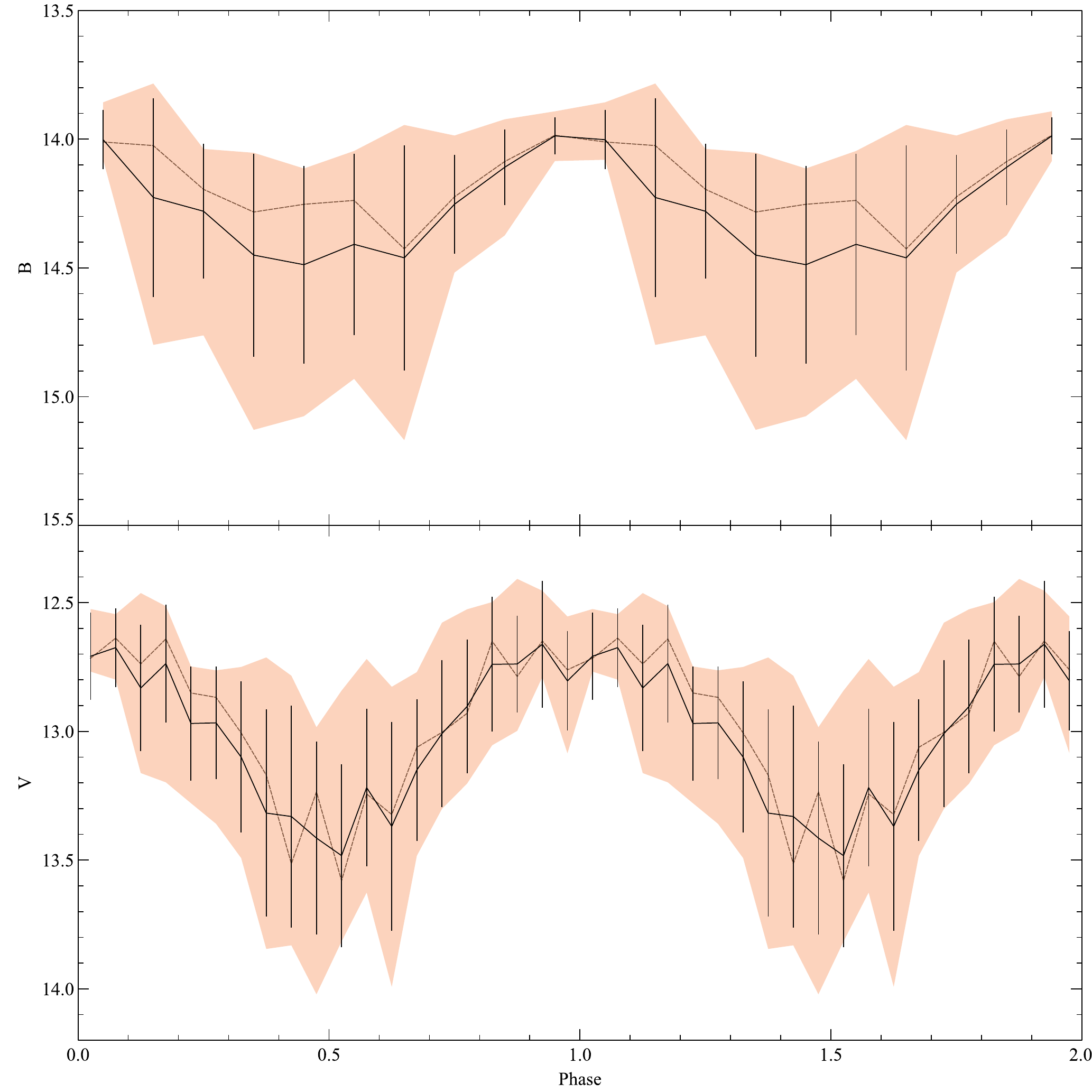}
\caption{Average light curves for $B$-band (top) and $V$-band (bottom) observations split across 10 bins per phase in $B$ and 20 bins per phase in $V$. The solid line shows the mean brightness in each bin and the dashed line shows the median. The standard deviation of each bin is shown by the error bars. The brightest 10 per cent and faintest 10 per cent of the data have been removed with the remaining 80 per cent shown in the shaded region.} 
\label{averagephot}
\end{figure}

It is clear that the AA Tau system varies considerably over time, so there will be a number of different parameter sets that provide the best fits to the data at different times. For this reason, we have averaged all of the available data and will compare this mean light curve with our models. We then compared the best fitting models for the averaged data with individual data sets to determine the set of parameters that best describes the system over the longest period of time. The average $B$ and $V$ light curves are presented in Fig.~\ref{averagephot}. $B$-band data are binned into 10 bins per phase and $V$-band data are binned into 20 bins per phase.
The solid line shows the mean magnitude in each bin along with the standard deviation. The median magnitude is given by the dotted line. The shaded region gives the range in magnitudes in which the median 80 per cent of data falls.

\begin{figure}
\centering
\includegraphics[width=87mm]{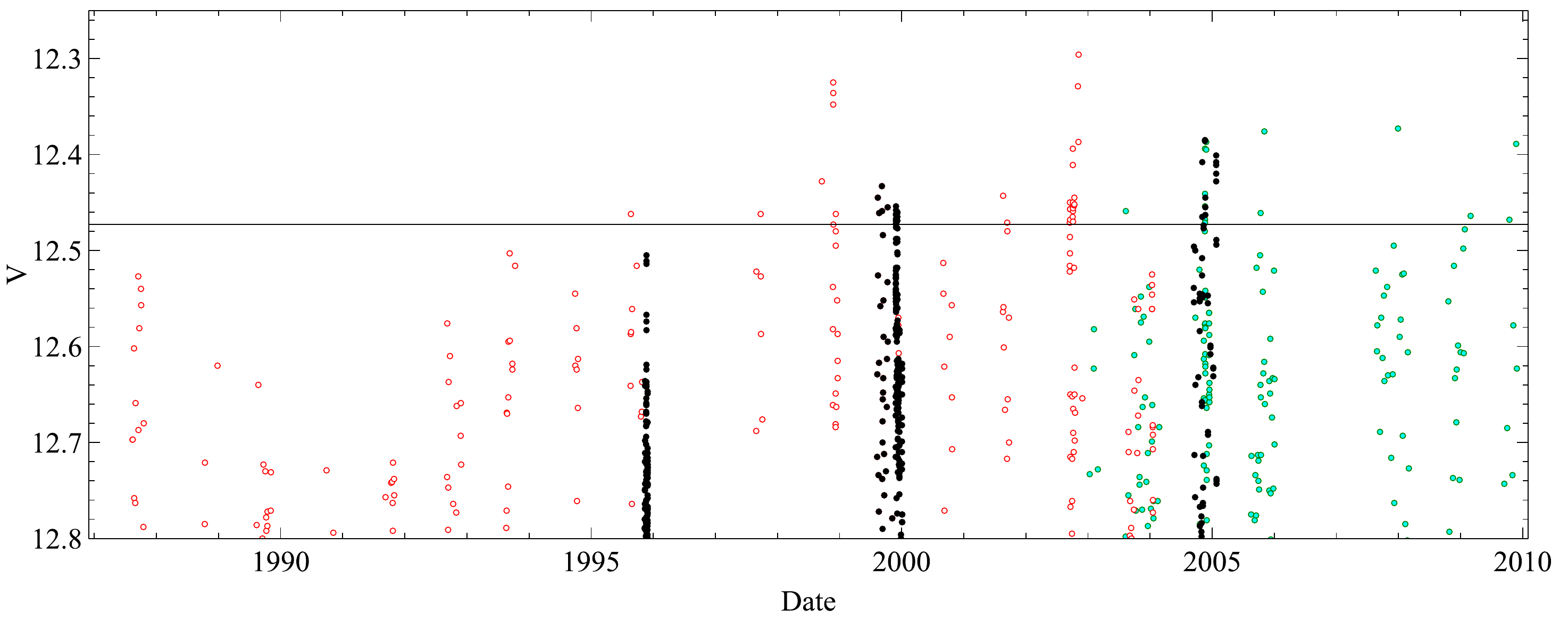}
\caption{Variations in brightness maxima for AA Tau from 1987 to 2010. Observations from \protect\citet{Grankin07} are shown in red, observations from B99, B03 and B07 in black, and observations from \protect\citetalias{ASAS} in green. The solid line shows the maximum out-of-occultation brightness computed by our models. A minimum in the observations is apparent around the early 1990s, with a maximum occurring 10 -- 15 years later. (Adapted from \protect\citealt{Bouvier13}.)} 
\label{variationsV}
\end{figure}

Comparison plots between the averaged observational data and our models are presented in Appendix \ref{appendixA}. One obvious feature is the offset between averaged observational data and modelled data in the $V$-band by about 0.2 mag. The data from B99 will contribute significantly to this due to the large number of observations obtained and the fairly consistent out-of-occultation $V$-band magnitude of 12.7. While there are other data sets that give even fainter \emph{average} out-of-occultation magnitudes, these have large amounts of scatter and, individually, fit the modelled brightnesses fairly well, as will be discussed. The maximum brightness has been shown to vary with time, with fluctuations around the 12.4-12.7 mag level over the last 25 years (Fig. \ref{variationsV}).

From Fig.~\ref{bouvierobs}, AA Tau is clearly bluer in 1995--96, which implies that the change in brightness is not inherent to the star itself. Since B99 $V$-band data are systematically fainter both in and out of occultation, the cause is also not likely to be a change in the warp structure. This points to an azimuthally symmetric density perturbation in the disc, obscuring the photosphere even when the warp is on the opposite side of the star. Since the only obscuring material included in the model is the warp itself, our expectation is that the model will systematically lie above the average $V$-band photometry, since we calculate the maximum brightness for a given parameter set. Disappearance of this perturbation would then cause the brightness to increase in $V$ back to the level calculated in the model, without significant change in the $B$-band since the density perturbation is restricted to the plane of the disc, thus not obscuring the hotspot.

The aspect ratio appears to be the parameter best constrained by observations. Synthetic light curves from a disc with $h_\mathrm{max}/r_{o}~\ge~0.33$ occult too much of the photosphere (see Figs \ref{o10h33} -- \ref{o20h33}). $B$-band models begin to fit reasonably well at $h_\mathrm{max}/r_{o} = 0.32$ at low dipole offsets but $V$-band models still show excessive occultation (Fig \ref{o10h32}). Further discussion will therefore be restricted to $h_\mathrm{max}/r_{o} =$~0.29, 0.30 and 0.31. As mentioned above, $B$-band fits will be selected not just according to the shape but also the brightness. A dipole offset of $40^\circ$ gives light curves which are too bright to match observations (Figs \ref{o40h29} and \ref{o40h30}). A $30^\circ$ offset does fall in the bright end of the range in the average $B$ photometry, but is still consistently brighter than the mean magnitude (Figs \ref{o30h29} and \ref{o30h30}). The dipole offset is therefore restricted to $10^\circ$ and $20^\circ$. The smallest magnetosphere size, with $r_i = 7.6$, does not produce sufficient occultation in the $B$-band for any of the models with $h_\mathrm{max}/r_{o}~\le~0.30$ (e.g. Fig \ref{o30h30}, dash-dotted line). While the fit is sufficient in $B$ for $h_\mathrm{max}/r_{o} = 0.31$, the occultation depth in $V$ is much greater than the mean magnitude (Fig \ref{o10h31}, bottom). The inner magnetosphere radius is therefore restricted to $r_i = 6.4$ and 5.2 R$_\star$.

\begin{figure}
\centering
\includegraphics[width=85mm]{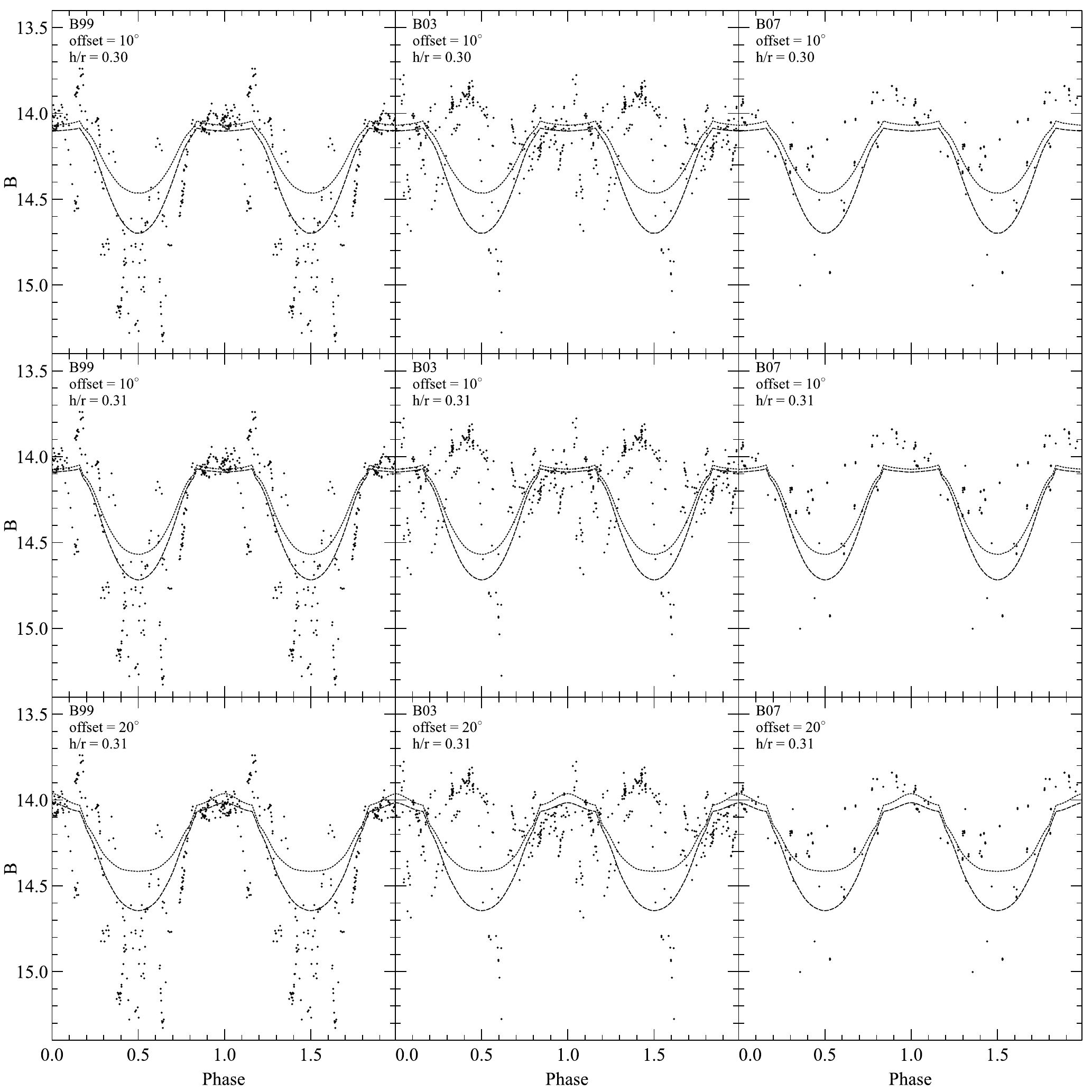}
\caption{Fits to $B$-band photometry. The data source and parameter set are labelled on the top left of each plot. Dashed lines show the largest magnetosphere size, $r_i = 5.2$ R$_\star$, and dotted lines show $r_i = 6.4$ R$_\star$.} 
\label{Bcomparison}
\end{figure}

Since the size of the magnetosphere has very little effect on $V$-band photometry, individual $B$-band data sets were used to determine the best magnetosphere size. The ranges determined above were considered, with the exception of $r_i = 6.4$ R$_\star$ where $\theta = 10^\circ$ or $20^\circ$ with $h_\mathrm{max}/r_{o} =$ 0.29, and where  $\theta = 20^\circ$ with $h_\mathrm{max}/r_{o} =$ 0.30, i.e. smaller magnetosphere with larger offset and smaller wall heights, since there is not enough occultation to produce the required luminosity dip. The $B$-band fits are given in Fig.~\ref{Bcomparison}. \citetalias{Bouvier03} data are difficult to fit due to the unusual shape of the light curve with two minima present in one phase. \citetalias{Bouvier07} data could fit either magnetosphere size, with evidence for both shallower and deeper minima than the models, although the smaller magnetosphere size does not fit the $\theta = 20^\circ$ model so well. Due to the large amplitude measured in \citetalias{Bouvier99}, the larger magnetosphere size provides the best fit. An inner magnetosphere radius of $r_i = 5.2$ R$_\star$ has therefore been selected.

\begin{figure}
\centering
\includegraphics[width=87mm]{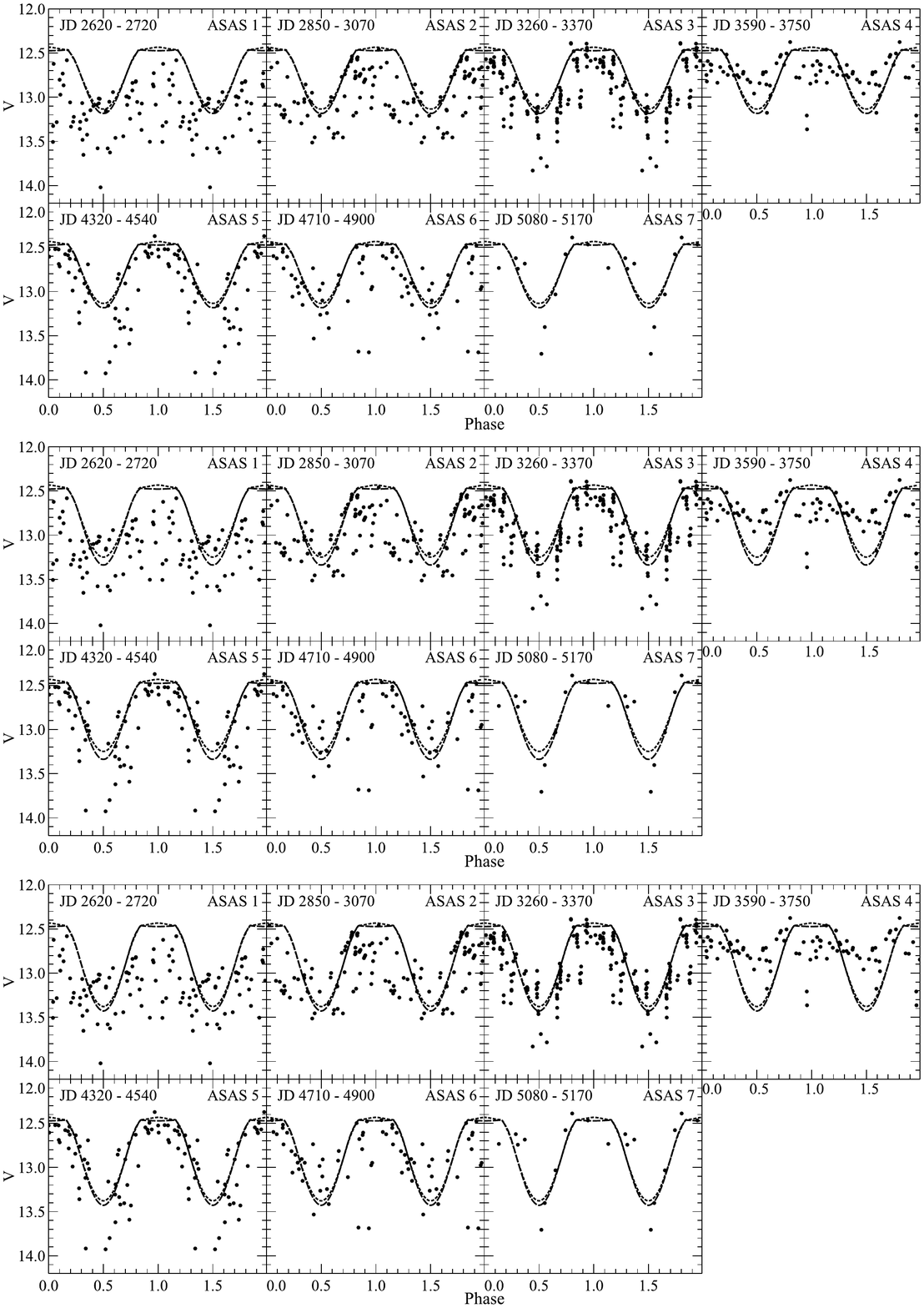}
\caption{Fits to \emph{ASAS} data for $h_\mathrm{max}/r_{o} =$ 0.29 (top), 0.30 (middle), and 0.31 (bottom), with $\theta = 10^\circ$ (dashed) and $\theta = 20^\circ$ (dotted). The date of each epoch is labelled on the top left of each plot.} 
\label{Vcomparison}
\end{figure}

\begin{figure}
\centering
\includegraphics[width=85mm]{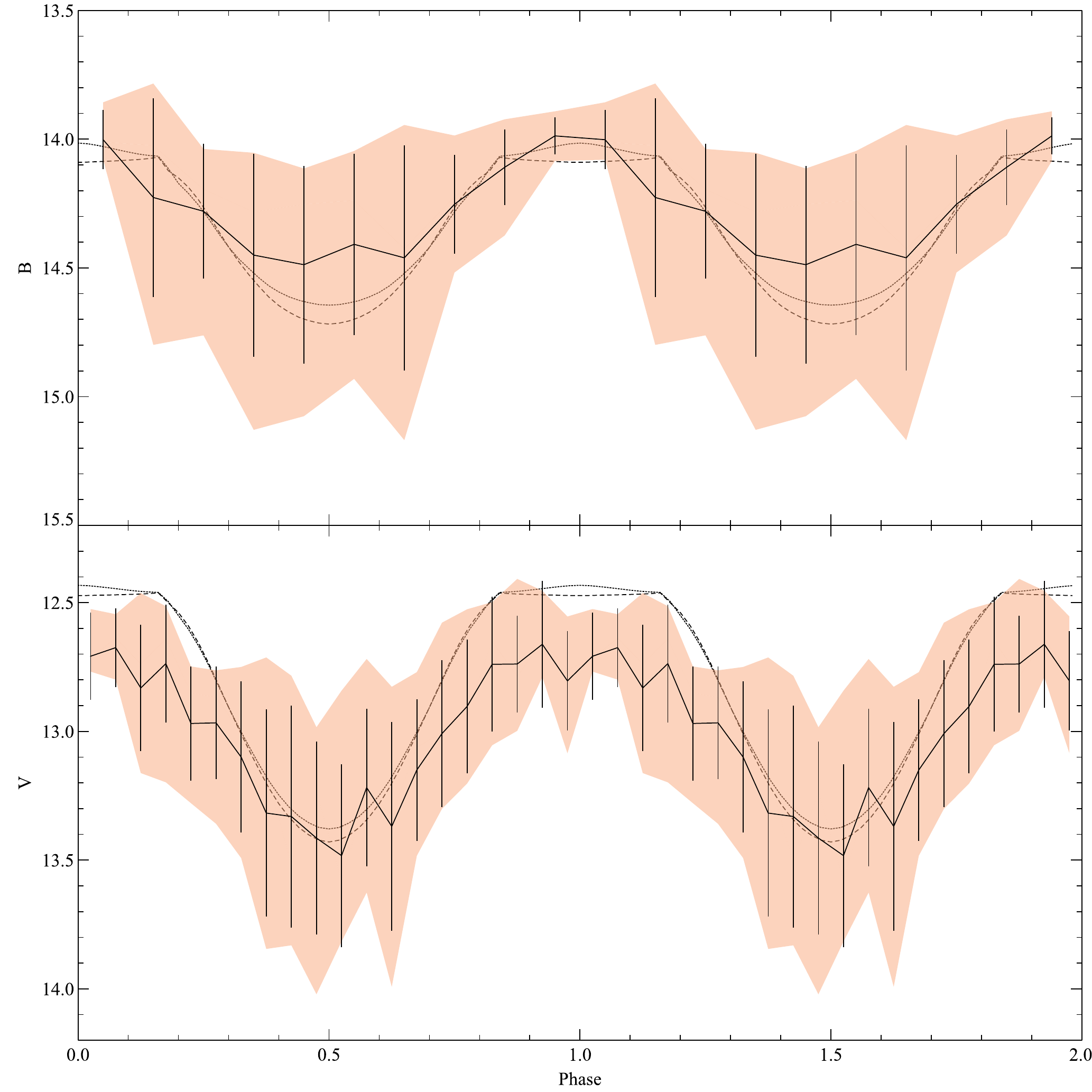}
\caption{Best fits to average photometric data from \citetalias{Bouvier99},  \citetalias{Bouvier03},  \citetalias{Bouvier07}, and  \citetalias{ASAS}. The solid line shows the mean brightness at each phase with the standard deviation given by the error bars. We found observations were best fitted by models with $\dot{M}_{acc} = 5\times10^{-9}$~M$_\odot$~yr$^{-1}$, $r_i = 5.2$ R$_\star$ and $h_\mathrm{max}/r_{o} =$ 0.31. The dashed line shows $\theta = 10^\circ$ and the dotted line shows $\theta = 20^\circ$.} 
\label{bestfits_photometry}
\end{figure}

$h_\mathrm{max}/r_{o}$ and $\theta$ have been selected by comparisons between models and \citetalias{ASAS} observations, since the \citetalias{Bouvier99} $V$-band data have a significant systematic offset, the \citetalias{Bouvier03} data contains two minima either side of $\varphi = 0.5$, and the \citetalias{Bouvier07} data were obtained at the same epoch as \citetalias{ASAS} \emph{3}. The remaining parameter sets are shown in Fig.~\ref{Vcomparison}. While models with $h_\mathrm{max}/r_{o} =$ 0.29 fit \citetalias{ASAS} \emph{6} well, and \citetalias{ASAS} \emph{2} when accounting for a slight $V$-band offset, they do not have a sufficient amplitude to fit the majority of epochs. Either of the remaining wall heights are acceptable, but $h_\mathrm{max}/r_{o} =$ 0.31 has been selected since it is the deeper of the two, better suiting \citetalias{ASAS} \emph{3} and \citetalias{ASAS} \emph{5} - two of the cleanest light curves. There is very little difference in models with $\theta = 10^\circ$ and $\theta = 20^\circ$. These best fits are shown in Fig.~\ref{bestfits_photometry}. We have selected $\theta = 10^\circ$ in preference to $\theta = 20^\circ$ on the basis of results from spectral line modelling (see Section \ref{mainspecresults}). 

In summary, the aspect ratio is fairly well constrained by photometry, with acceptable fits given by $h_{max}/r_{o}=0.29$ -- 0.31 and a best fit given by $h_{max}/r_{o}=0.31$. The inner magnetosphere radius and dipole offset are less well constrained. We have found $r_i \sim 5.2$ -- 6.4 R$_\star$, with 5.2 R$_\star$ selected on the basis that it gives a deeper minimum during occultation, better fitting some of the individual data sets. Dipole offset values of $\theta < 30^\circ$ provide the best fits, but $\theta$ cannot be constrained beyond this with photometry alone.

\section{Spectroscopy}
\label{sec:spectroscopy}
In a similar method to the photometric study described above, a number of synthetic spectra for AA Tau have been modelled over various parameters using the radiative transfer code {\sc torus}, where results were compared with observed hydrogen spectra obtained by \citetalias{Bouvier07}. Observational results are given in \ref{sec:specObs} and the model is described in Section \ref{sec:modelCalcs}. Synthetic spectra for AA Tau are presented in Section \ref{sec:AATauSpec}.

\subsection{Observational spectroscopy}
\label{sec:specObs}

\begin{figure*}
\centering
\includegraphics[width=150mm]{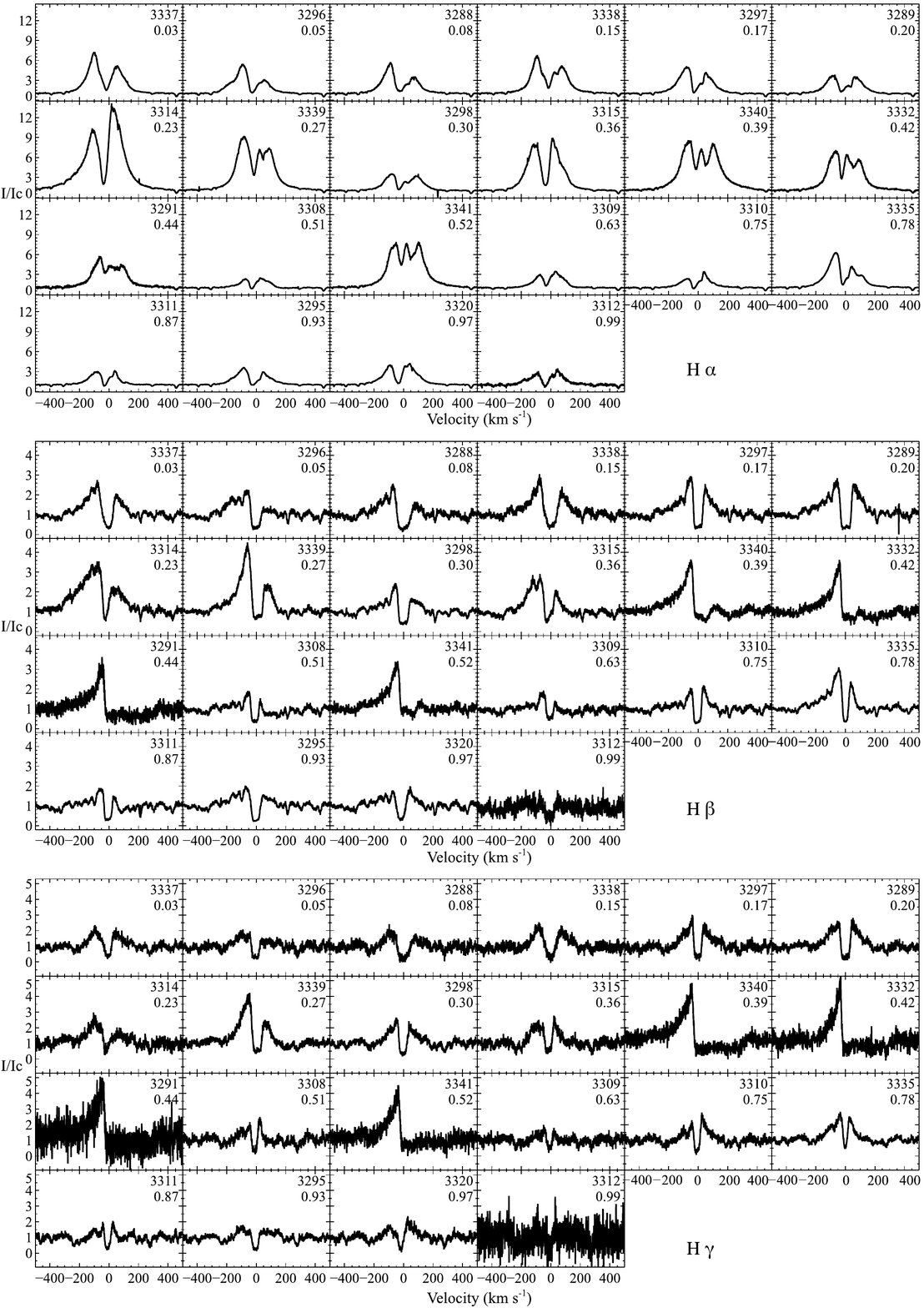}
\caption{Emission line profiles for H{\galpha} (top), H{\gbeta} (middle), and H{\ggamma} (bottom), from \protect\citet{Bouvier07}. Profiles are ordered by rotational phase. JD~$-$~2450000 is given on the top right of each profile, with phase underneath.}
\label{allspec}
\end{figure*}

Emission line profiles for H{\galpha}, H{\gbeta} and H{\ggamma}, from \citetalias{Bouvier07}, were taken between October and December 2004 (JD~2453288 -- 341) at the ESO La Silla 3.6~m telescope with the HARPS high-resolution echelle spectrograph. 22  spectra for each line were obtained in total, covering the 3800 -- 6900~\AA\ spectral domain at a spectral resolution of $\lambda/\Delta\lambda \approx 115\,000$ with a signal-to-noise ratio between 10 and 30 at 600~nm. The line profiles have been phased over an 8.22 d period, as with the photometry. They were rectified by manually defining line-free continuum bins, performing a polynomial fit to the continuum level, and dividing each line profile by the resulting continuum. Rectified line profiles are shown in Fig.~\ref{allspec}. We have focussed on fitting H{\gbeta} which (at some phases) displays the classical inverse P Cygni (IPC) absorption which is associated with the magnetospheric accretion paradigm.

\begin{figure}
\centering
\includegraphics[width=85mm]{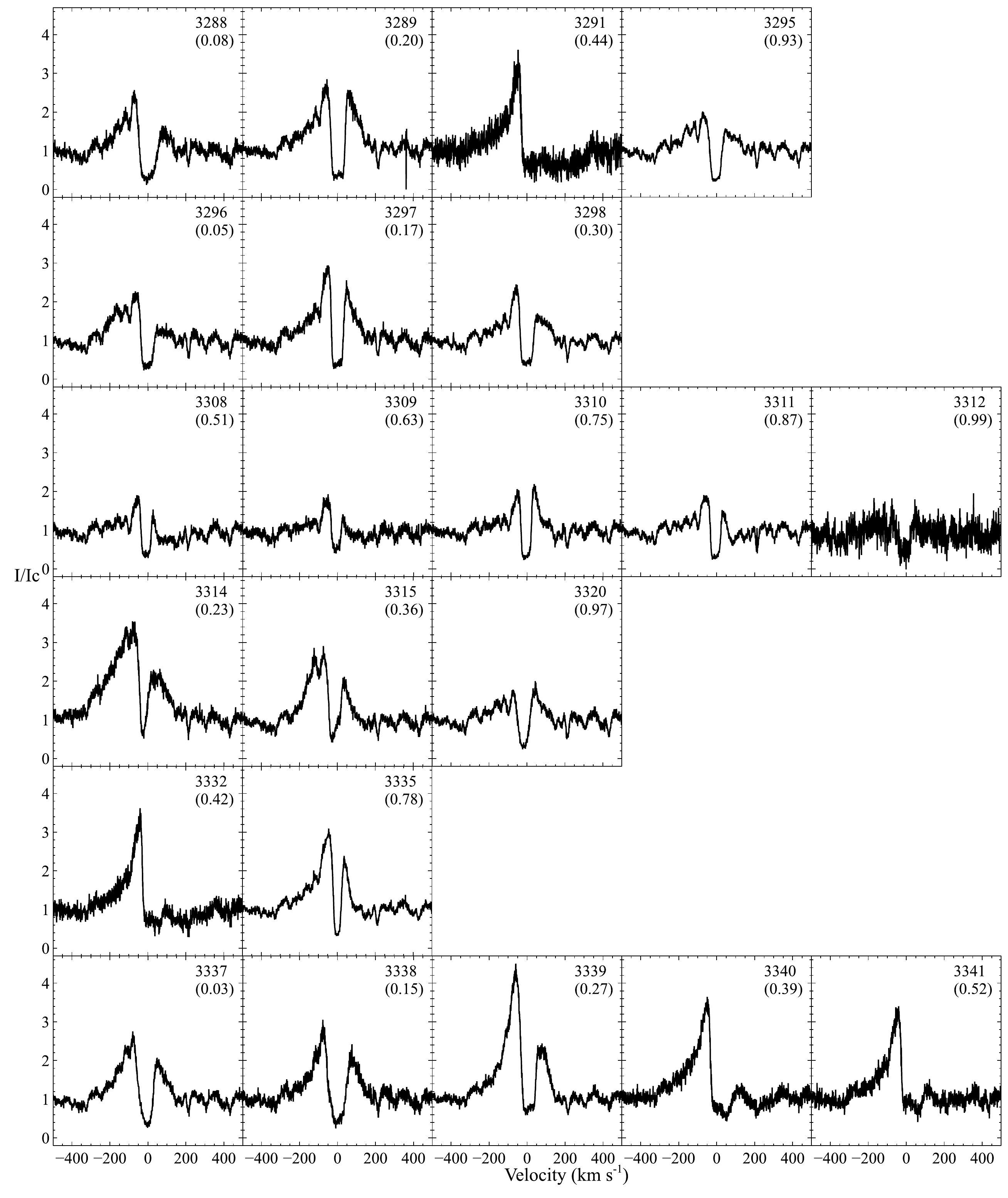}
\caption{Observational spectra for H\gbeta, ordered by date. Spectra obtained during one rotation are plotted in each row. JD~$-$~2450000 is given in the top right and phase underneath.}
\label{dateordered}
\end{figure}

As with photometry, the shape and intensity of emission line profiles vary over short time-scales, with variations often apparent between profiles separated by just one or two rotations. The phase ordered spectra will therefore appear to vary somewhat erratically from one phase to the next. However, when ordered by date and grouped into individual rotations, variations between adjacent spectra are reduced and more of a trend is apparent during an individual rotation (e.g. see Fig.~\ref{dateordered} for H\gbeta). An interpretation of the spectral features seen in these profiles is given in \citetalias{Bouvier07}, but in summary:
\begin{itemize}
\item[--] A deep, blueshifted central absorption is always present, consistent with an outflow; 
\item[--] High-velocity redshifted absorption for phases around $\varphi=0.5$, i.e. during photometric minimum, with the result that H{\galpha} profiles have a triple-peaked appearance around this phase, whereas in H{\gbeta} and H{\ggamma} profiles the red wing is suppressed completely. In some instances the absorption falls below the continuum level in an IPC profile shape. This redward absorption is consistent with a line of sight down the main accretion funnel flow;
\item[--] The redshifted absorption component is not as pronounced in cycles with a shallow photometric minimum (e.g. JD 3308), consistent with a low accretion rate and thus a small (perhaps negligible) disc warp.
\end{itemize}

Each of these features have been found in numerous observations of other CTTs to varying degrees, with H{\galpha} typically demonstrating blueshifted central absorption more so than higher lines, but with higher lines typically showing IPC shapes \citep[e.g.][]{Edwards94,Alencar00,Folha01}.

\subsection{AA Tau Synthetic Spectra}
\label{sec:AATauSpec}
As with photometry, different parameter sets will give the best fits at different times due to the variability of AA Tau's magnetic field.

Each model produced line profiles from 20 different rotational phases. Natural, van der Waals, and Stark broadening for H{\galpha} was included using the parameters listed in Table \ref{tab:windparams}, but no broadening was included for H{\gbeta} and H{\ggamma} lines since broadening is less important here.
The best fitting parameters obtained in the photometric study  ($\dot{M}_{acc}=5\times10^{-9}$~M$_{\odot}$~yr$^{-1}$, $r_i=5.2$~R$_{\star}$, $h_{max}/r_{o}=0.31$) have been kept constant. The maximum temperature in the magnetosphere $T_{mag}$, wind acceleration parameter $\beta$, mass-loss from the disc wind $\dot{M}_{wind}$, and wind temperature $T_{wind}$ were varied over dipole offsets of $\theta=10^{\circ}$ and $20^{\circ}$. Values for $\dot{M}_{wind}$ were chosen such as to give $\dot{M}_{wind}/\dot{M}_{acc}$ values of 0.05, 0.10, 0.15 and 0.20 to study the parameter space around the canonical value of 0.1. Magnetosphere temperature $T_{mag}$ was modelled at  8000, 8500 and 9000~K and wind temperatures $T_{wind}$ at 7000, 8000 and 9000~K. $\beta$~values were modelled at 0.5, 1.0 and 2.0. The full range of parameter space is presented in Table~\ref{tab:windparams}.

\begin{table}
\caption{\label{tab:windparams}\small{Parameters used for the spectroscopic study, in addition to the stellar parameters given in Table \ref{tab:params}. The disc wind parameters $r_{wi}$ to $f$ are taken from \protect\citet{Kurosawa06}. The broadening parameters are taken from \protect\citet{LuttermoserJohnson92} and are for H{\galpha} only. Model parameters  $T_{mag}$, $\beta$, $\dot{M}_{wind}$ and $T_{wind}$ have been fitted using the values listed below.}}
\begin{center}
\begin{tabular}{| l | l | r |}\hline
Parameter & Description & Value\\ \hline
$r_{wi}$ & Inner disc wind radius & 1.0 r$_\mathrm{o}$\\
$r_{wo}$ & Outer disc wind radius & 35.6 r$_\mathrm{o}$\\
$d$ & Position of wind source points & 7.33 r$_\mathrm{wi}$\\
$\alpha$ & Exponent in mass-loss rate per unit area & 0.76\\
$\gamma$ & Exponent in disc temperature power law & -1.15\\
$R_s$ & Effective acceleration length & 8.8 r$_\mathrm{wi}$\\
$f$ & Scaling on the terminal velocity & 2.0\\\\
$C_{rad}$ & Natural broadening parameter & $6.4\times10^{-4}$ \AA\\
$C_{vdW}$ & van der Waals broadening parameter & $4.4\times10^{-4}$ \AA\\
$C_{Stark}$ & Stark broadening parameter & $1.17\times10^{-3}$ \AA\\
$\lambda_{H\alpha}$ & H{\galpha} wavelength & 6562.8 \AA\\
$\lambda_{H\beta}$ & H{\gbeta} wavelength & 4860.9 \AA\\
$\lambda_{H\gamma}$ & H{\ggamma} wavelength & 4340.0 \AA\\\\
$T_{mag}$ & Magnetosphere temperature & 8000~K\\
&&8500~K \\
&&9000~K \\
$\beta$ & Wind acceleration parameter & 0.5\\
&&1.0\\
&&2.0\\
$\frac{\dot{M}_{wind}}{\dot{M}_{acc}}$ & Ratio of mass-loss rate& 0.05 -- 0.15\\
&to mass accretion rate&\\
$T_{wind}$ & Disc wind temperature & 7000 -- 9000~K\\\hline
\end{tabular}
\end{center}
\end{table}

\subsubsection{Magnetosphere only}

\begin{figure*}
\centering
\includegraphics[width=130mm]{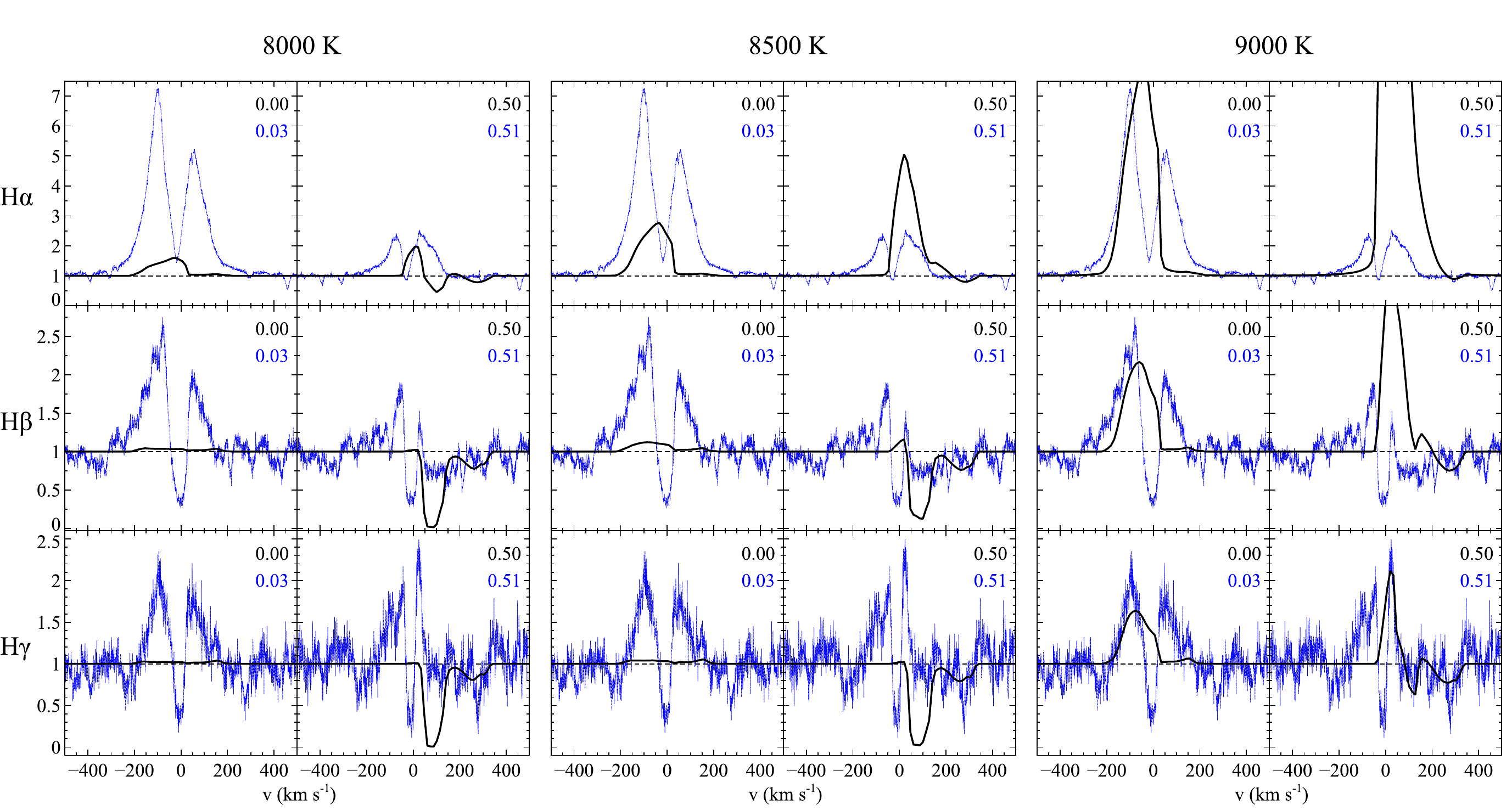}
\caption{Magnetosphere-only models for H{\galpha}, H{\gbeta} and H{\ggamma} at a range of temperatures at $\varphi=0.0$ and $\varphi=0.5$. Observed profiles are given in blue, and the continuum level is shown by the dashed line.}
\label{magonly}
\end{figure*}

Spectra for magnetosphere-only models have been calculated to address the question of whether observed profiles can be fitted by a model with no disc wind. A study of H{\galpha} line profiles by \citet{Kurosawa06} found that magnetosphere-only models were not able to reproduce all of the properties observed in CTT profiles, although the accretion in their model was axisymmetric with no dipole offset. Here we use our non-axisymmetric magnetosphere-only models, including a disc warp, to draw comparisons between observations and synthetic spectra for AA Tau. A selection of line profiles are presented in Fig.~\ref{magonly} for H{\galpha}, H{\gbeta} and H{\ggamma} at each magnetosphere temperature. Models are shown at phases of 0.0 (out of occultation) and 0.5 (during occultation) with $\theta = 10^\circ$.

There are features of the magnetosphere-only model that are found in observed profiles such as redshifted absorption components at $\varphi=0.5$, as expected for observations down the accretion stream. Another feature is the blueshifted emission at $\varphi=0.0$, when the accretion stream in the observable hemisphere is at the back of the star relative to the line of sight. However, it is quite clear that the slightly blueshifted deep absorption apparent in observations is not reproduced by any of these magnetosphere-only models. We have included a disc wind in the following models in an attempt to resolve this discrepancy. 

\subsubsection{Magnetosphere and wind}
\label{mainspecresults}

\begin{figure*}
\centering
\includegraphics[width=180mm]{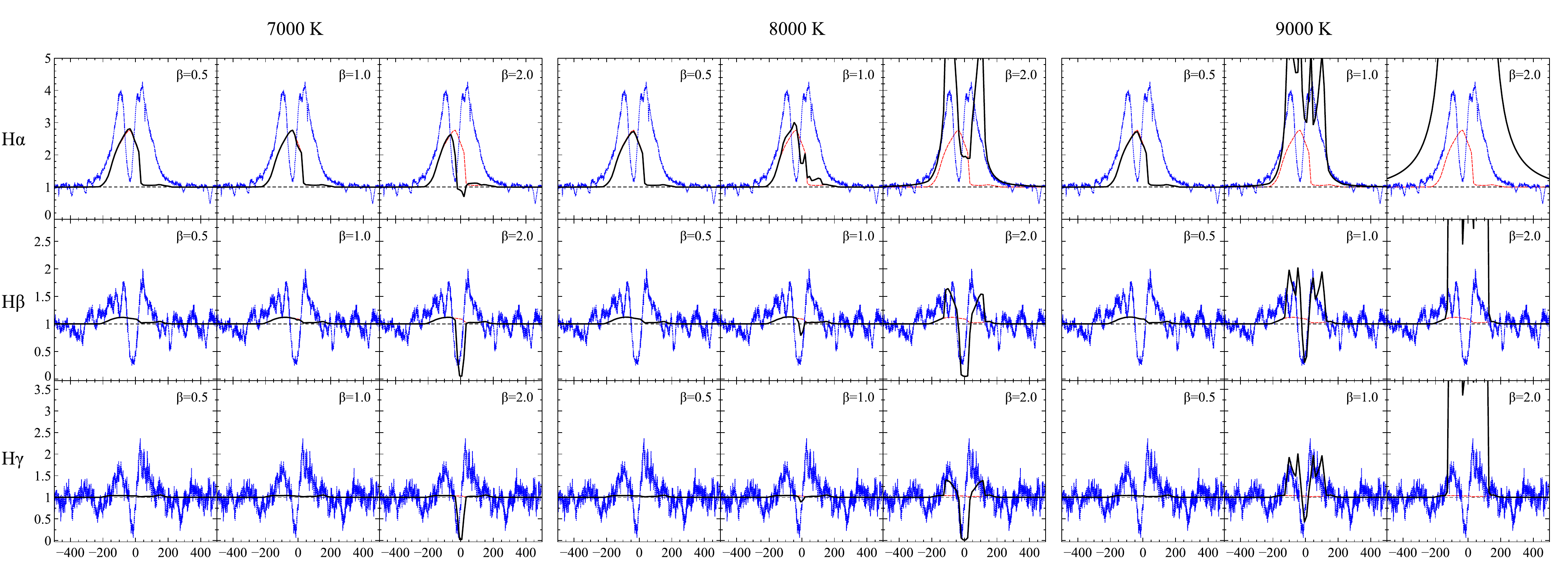}
\caption{Magnetosphere with wind models for H{\galpha}, H{\gbeta} and H{\ggamma} at $\varphi=0.0$, with $\theta=10^\circ$. Wind temperature and the acceleration parameter are varied with $T_{wind}=$ 7000, 8000 and 9000~K, and $\beta =$ 0.5, 1.0 and 2.0. Our line profiles are shown in black, observed profiles are shown in blue, and the magnetosphere contribution is given in red.}
\label{hybridmodels}
\end{figure*}

We have included a disc wind as described in Section \ref{sec:modelCalcs} \citep[see also][]{Kurosawa06} with wind parameters as given in Table \ref{tab:windparams}. We first studied the effect of wind temperature $T_{wind}$ and acceleration $\beta$, with results given in Fig.~\ref{hybridmodels}. Our line profiles are given in black with observations in blue. Magnetosphere contributions are shown in red. The central deep absorption is present in our models for hotter, more slowly accelerating winds when the magnetosphere contributes less to the profile shape relative to the wind.

Very little emission is present for models where $T_{wind}=$ 7000~K, expect for H{\galpha} profiles, although the deep absorption component appears when $\beta=2.0$. For winds at 8000~K the modelled profile shape matches the observed H{\gbeta} profile reasonably well for $\beta=2.0$, although the model is a little weak for H{\ggamma} and strong for H{\galpha}. A similar result is found for $T_{wind}=$ 9000~K with $\beta=1.0$. There is a degeneracy for wind temperatures between 8000~K and 9000~K and acceleration parameters between 1.0 and 2.0 where a number of different values will also provide sufficient fits to the observations. Physically this is a result from the degeneracy between density and temperature. The density of the wind is coupled to the acceleration parameter and the mass-loss rate, with temperature contributions defined by $T_{mag}$ and $T_{wind}$. We have selected a wind temperature of $T_{wind}=$ 8000~K and $\beta=2.0$ for our parameter fitting for the disc wind, but even by eliminating $\beta$ as a free parameter there will still be a degeneracy in density from the mass outflow rate, where the greater intensity in line profiles with an increased mass outflow rate can be reduced again by decreasing one or both of the temperature parameters. For this reason, we are unable to extract a definitive best fitting parameter set to describe the disc wind, especially since we are restricted to hydrogen recombination lines. It may be possible to break this degeneracy with additional lines such as He\,{\sc i} \citep*[e.g.][]{Kwan07,Kurosawa11}.

\begin{figure}
\centering
\includegraphics[width=85mm]{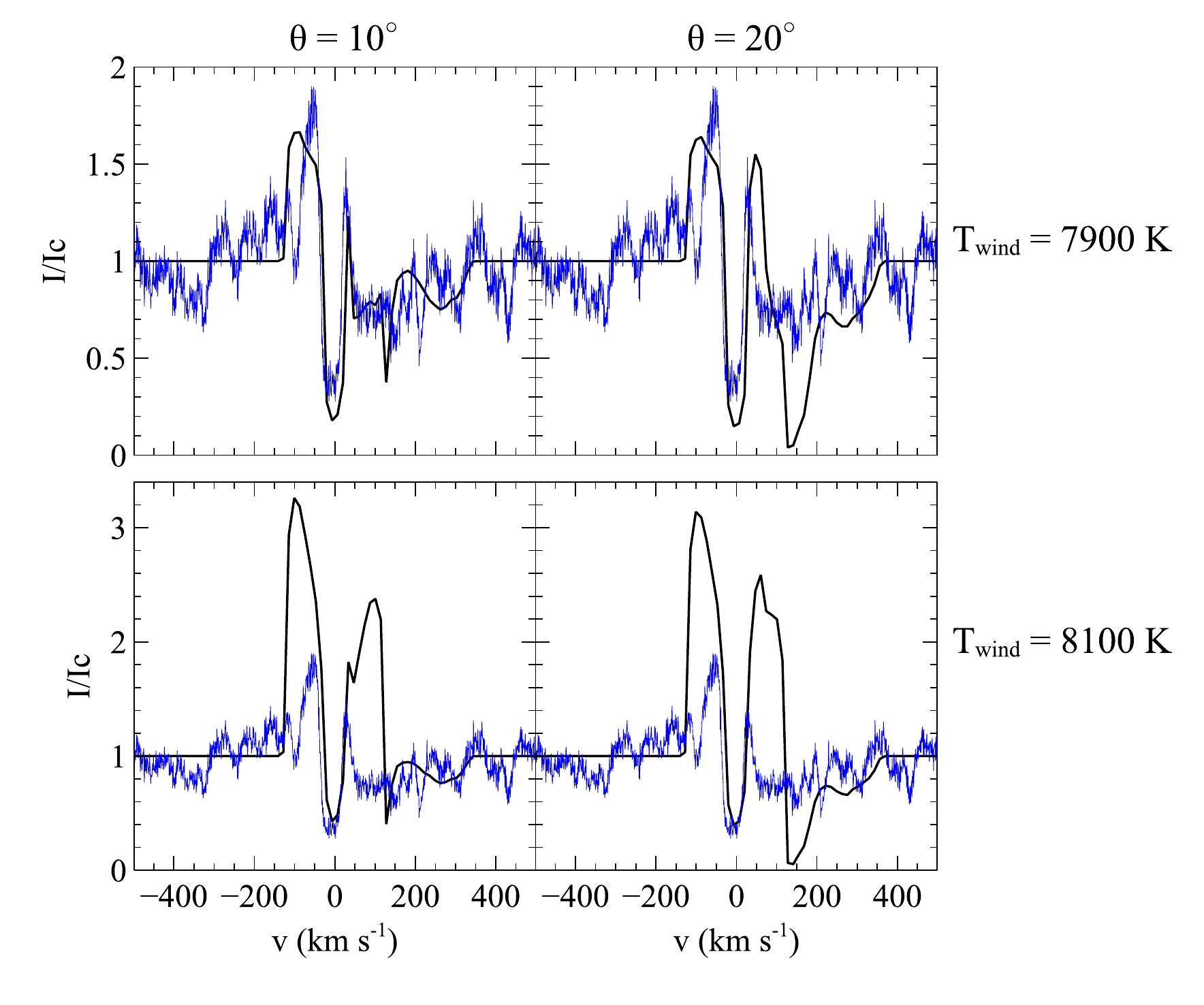}
\caption{Comparisons between observed spectra and models for H{\gbeta} with $\theta=10^\circ$ (left) and $\theta=20^\circ$ (right). $\dot{M}_{wind}/\dot{M}_{acc}$ = 0.1 and $T_{mag} = 8500$ K in each case. $T_{wind}$ is modelled at 7900~K (top) and 8100~K (bottom).}
\label{offsetspec}
\end{figure}

A more precise dipole offset angle appears to be better constrained by spectroscopy than photometry, having already been restricted to about $10^\circ$ -- $20^\circ$. There is very little difference between the two out of occultation, as to be expected, but differences do become apparent during occultation. Models at $\varphi = 0.5$ for H{\gbeta} are shown in Fig.~\ref{offsetspec} to demonstrate this, with $T_{wind}$ and $\beta$ as above. There is high-velocity redshifted absorption present for both $\theta = 10^\circ$ and $\theta = 20^\circ$, but the absorption is stronger for the $\theta = 20^\circ$ cases. There is also a difference in the relative emission strengths of the blue and red peaks, with a greater blue-to-red ratio in the $\theta = 10^\circ$ cases. Since observations show only slight high-velocity redshifted absorption and stronger emission in the blue peak than the red (with the red completely damped in some cases), we have chosen a dipole offset of $\theta = 10^\circ$.

\begin{figure}
\centering
\includegraphics[width=85mm]{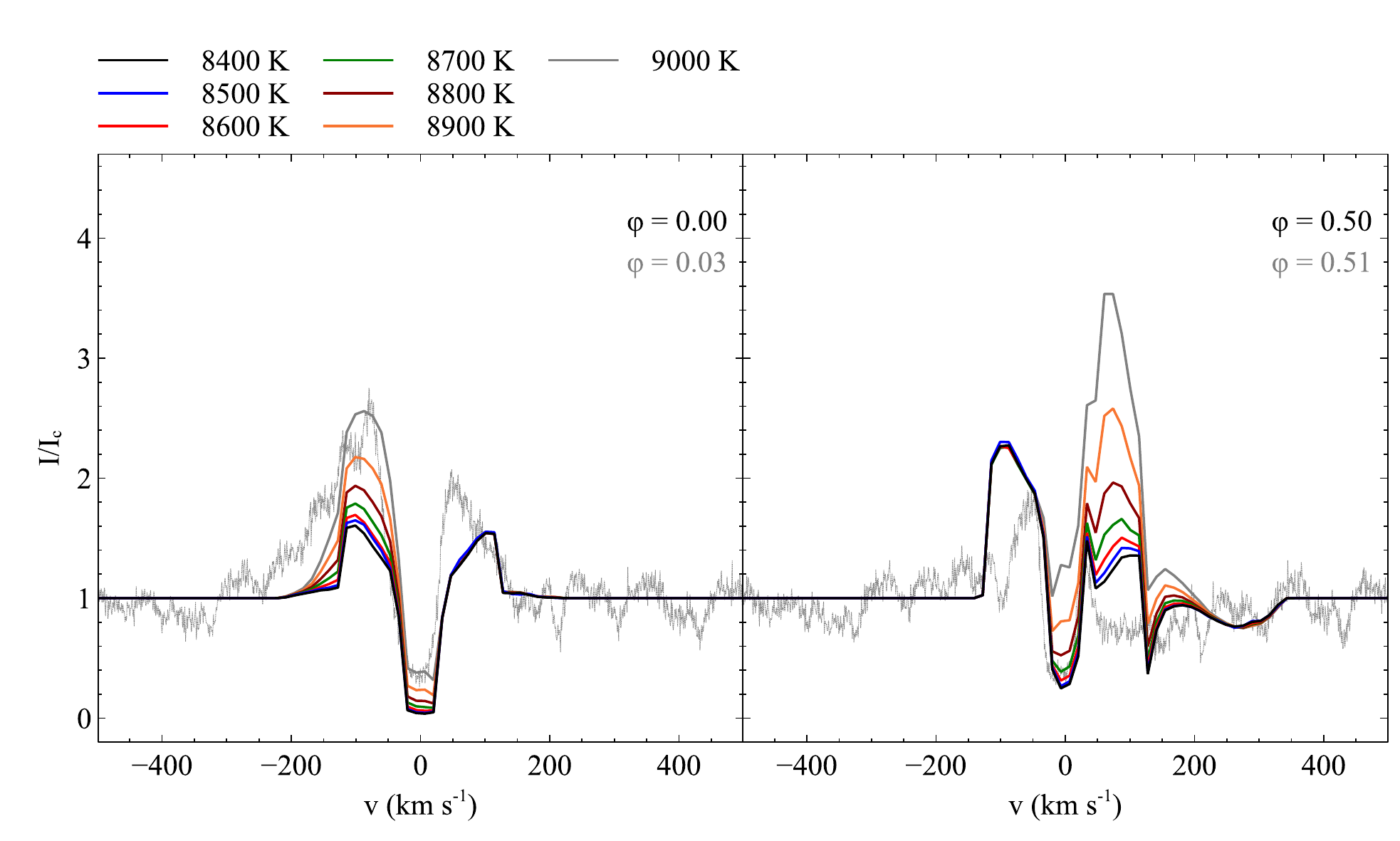}
\caption{Variation of the H{\gbeta} profile shape with different $T_{mag}$ values. $\varphi=0.0$ is shown on the left and $\varphi=0.5$ on the right. Observed profiles are in grey for comparison. All profiles are calculated with $\dot{M}_{wind}/\dot{M}_{acc}=$ 0.1 and $T_{wind}=$ 8000~K.}
\label{compareTmag}
\end{figure}

For the $\beta$ and $T_{wind}$ values determined above there is very little magnetosphere contribution to the line profile, so we have neglected further study of magnetosphere temperatures below 8400~K. We have assigned a maximum magnetosphere temperature of 9000~K, since this produces an H{\gbeta} profile of comparable intensity to observations at $\varphi=0.0$, and exceeds the observed intensity of some of the observed profiles around $\varphi=0.5$, without any consideration for the addition of a disc wind. The effect of the magnetosphere temperature on H{\gbeta} profiles in the range of 8400~K and 9000~K is shown in Fig.~\ref{compareTmag}. Note that the only considerable change between models with different magnetosphere temperatures at $\varphi=$ 0.0 is the strength of the blueward emission due to the visibility of part of the accretion stream behind the star. There is also a slight increase in the broadening of the blue wing at the highest temperatures. The same variations in intensity occur at $\varphi=$ 0.5 for redward emission, where the effect is more pronounced since the entire accretion stream is in view.  In general the blue peak is the primary in these observations, pointing towards a magnetosphere temperature of 8500~K which maintains the largest achievable difference in intensity between the primary and secondary peaks. However, higher temperatures are required to achieve sufficient broadening, in which case the wind must be able to dampen the higher-intensity red emission to match observations during occultation.

\begin{figure}
\centering
\includegraphics[width=85mm]{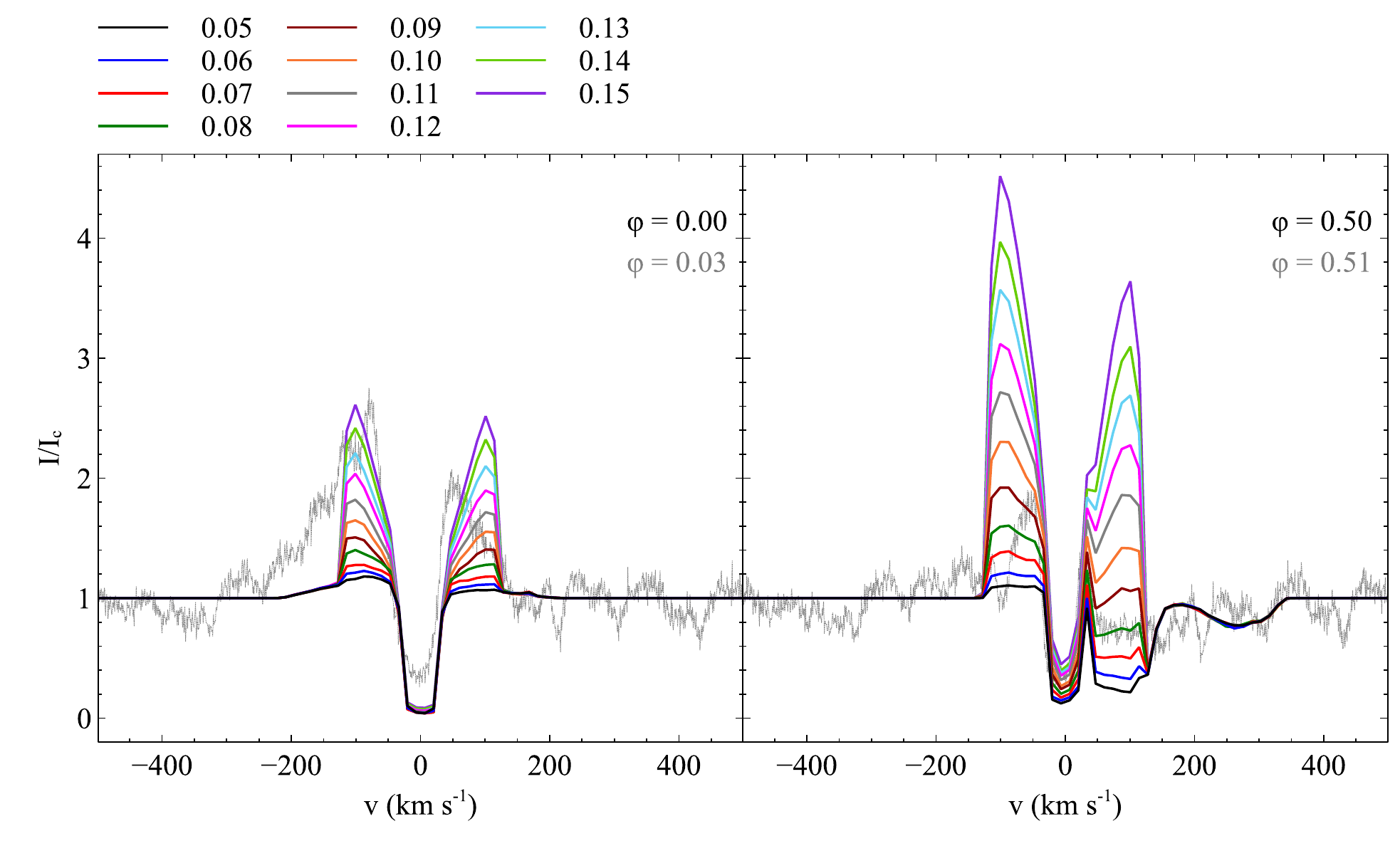}
\caption{Variation of the H{\gbeta} profile shape with different $\dot{M}_{wind}/\dot{M}_{acc}$ values. $\varphi=0.0$ is shown on the left and $\varphi=0.5$ on the right. Observed profiles are in grey for comparison. All profiles are calculated with $T_{mag}=$ 8500~K and $T_{wind}=$ 8000~K.}
\label{compareRatio}
\end{figure}

$T_{wind}$ is restricted to 8000 $\pm$ 500~K for $\beta =$ 2.0 and $T_{mag}=$ 8500~K, and $\dot{M}_{wind}/\dot{M}_{acc}$ has been varied between 0.05 and 0.15, in line with the canonical value of 0.1, by keeping the accretion rate constant at $5\times10^{-9}$~M$_\odot$~yr$^{-1}$ and varying the mass-loss rate.
The effect of varying $\dot{M}_{wind}/\dot{M}_{acc}$ is shown in Fig.~\ref{compareRatio}. There is little variation between profiles with ratios separated by 0.01 out of occultation. Variation between profiles is greater during occultation, with the red wing switching from absorption to emission at $\dot{M}_{wind}/\dot{M}_{acc}=0.09$.

\begin{figure}
\centering
\includegraphics[width=85mm]{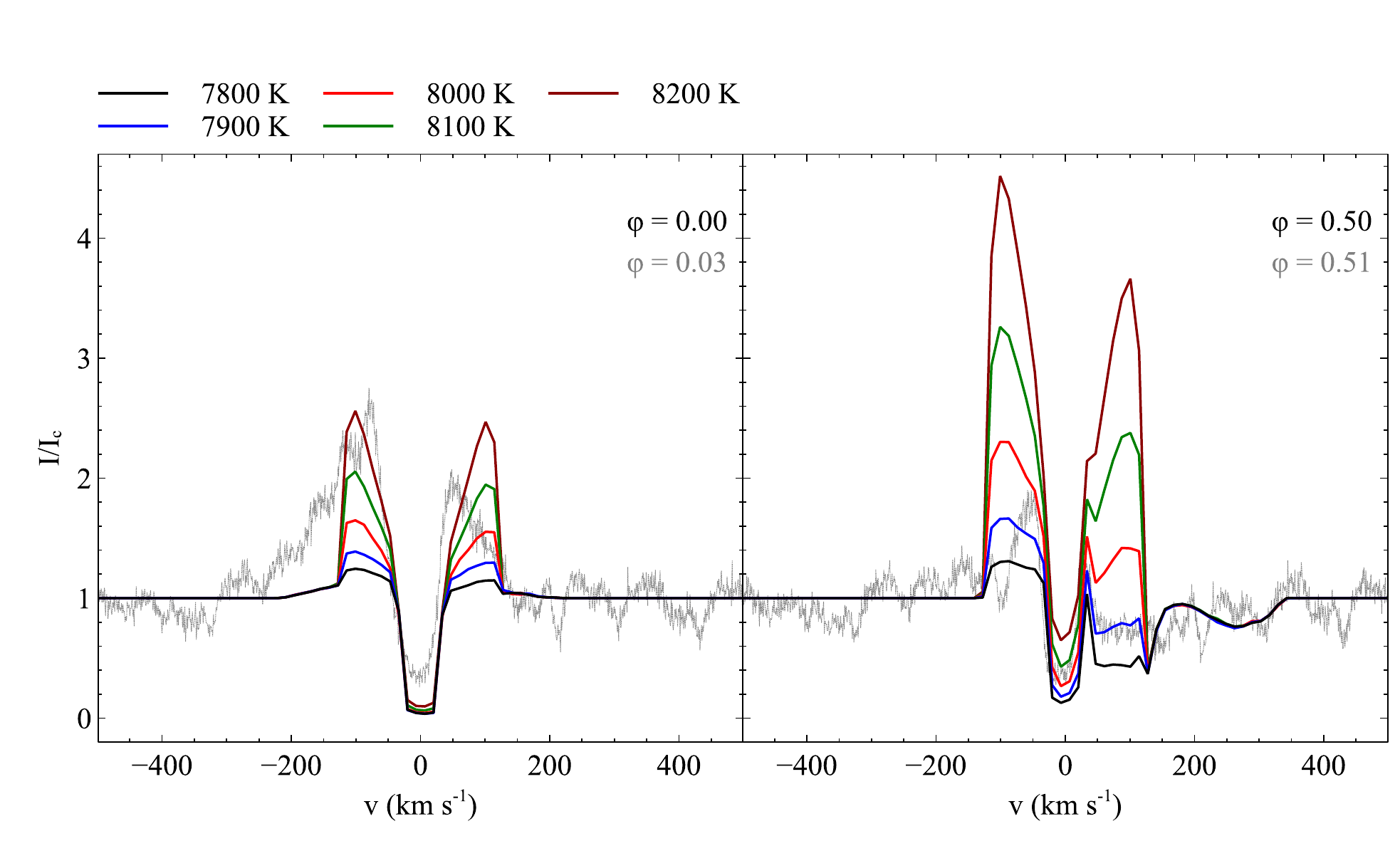}
\caption{Variation of the H{\gbeta} profile shape with different $T_{wind}$ values. $\varphi=0.0$ is shown on the left and $\varphi=0.5$ on the right. Observed profiles are in grey for comparison. All profiles are calculated with $\dot{M}_{wind}/\dot{M}_{acc}=$ 0.1 and $T_{mag}=$ 8500~K.}
\label{compareTwind}
\end{figure}

While models with higher mass-loss rates fit the observations better out of occultation, ratios around 0.1 fit better during occultation, with higher mass-loss rates giving excess emission. The variation in profile shape with different values for $T_{wind}$ is plotted in Fig.~\ref{compareTwind}. Similarly, different values fit the observed profiles better at different phases, with $\sim$ 8100--8200~K best describing the profile at $\varphi=0.0$ and $\sim$ 7900--8000~K best describing the profile at $\varphi=0.5$.

\begin{figure*}
\centering
\includegraphics[width=105mm]{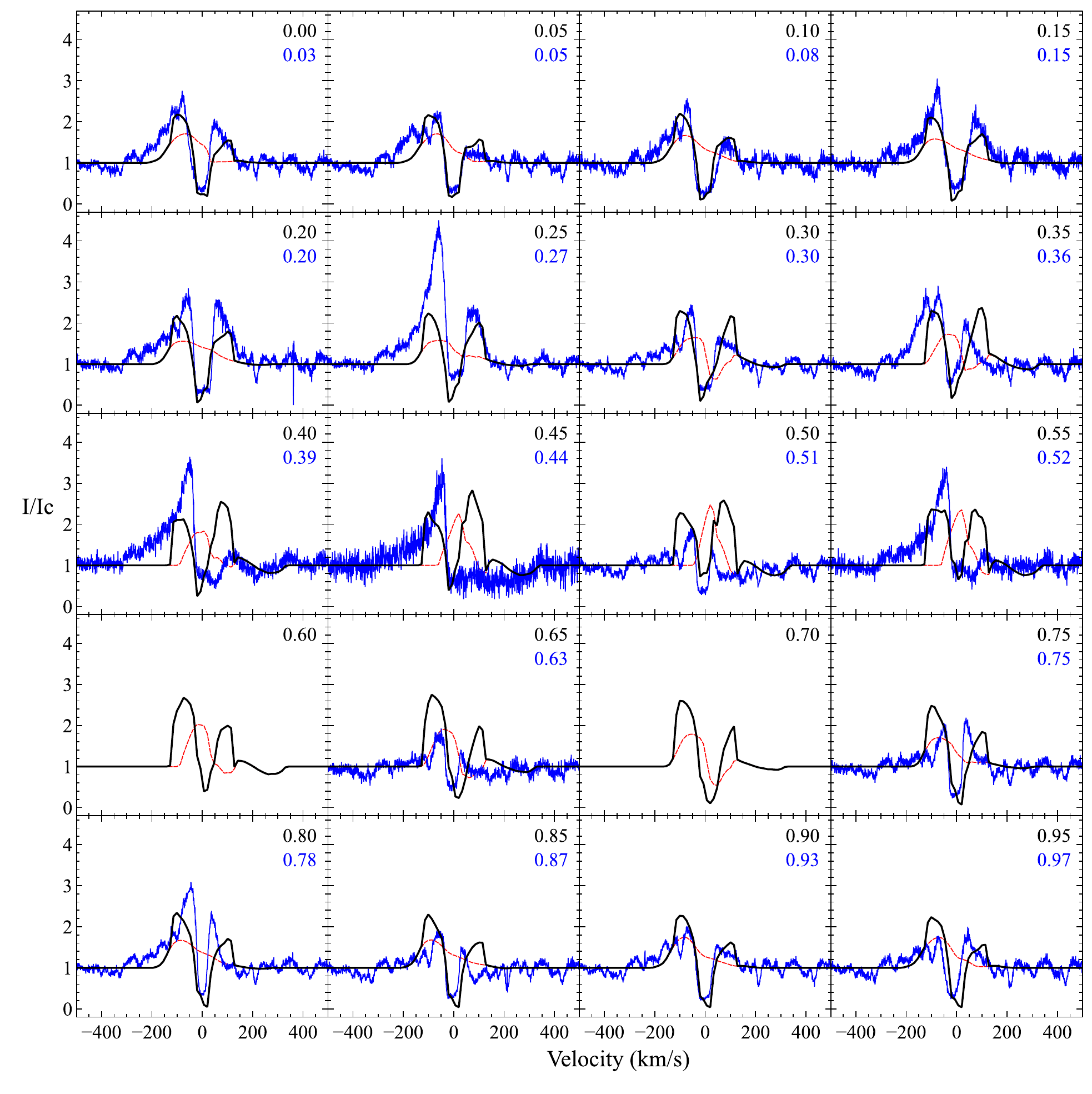}
\caption{Best fitting models for H{\gbeta} profiles out of occultation, with $\dot{M}_{wind}/\dot{M}_{acc}=0.1$, $T_{mag}=$ 8900~K and $T_{wind}=$ 8000~K. The magnetosphere contribution is shown by the red dotted line, and the magnetosphere-plus-wind profile is shown by the black solid line.}
\label{linesfit_outofocc}
\end{figure*}

\begin{figure*}
\centering
\includegraphics[width=105mm]{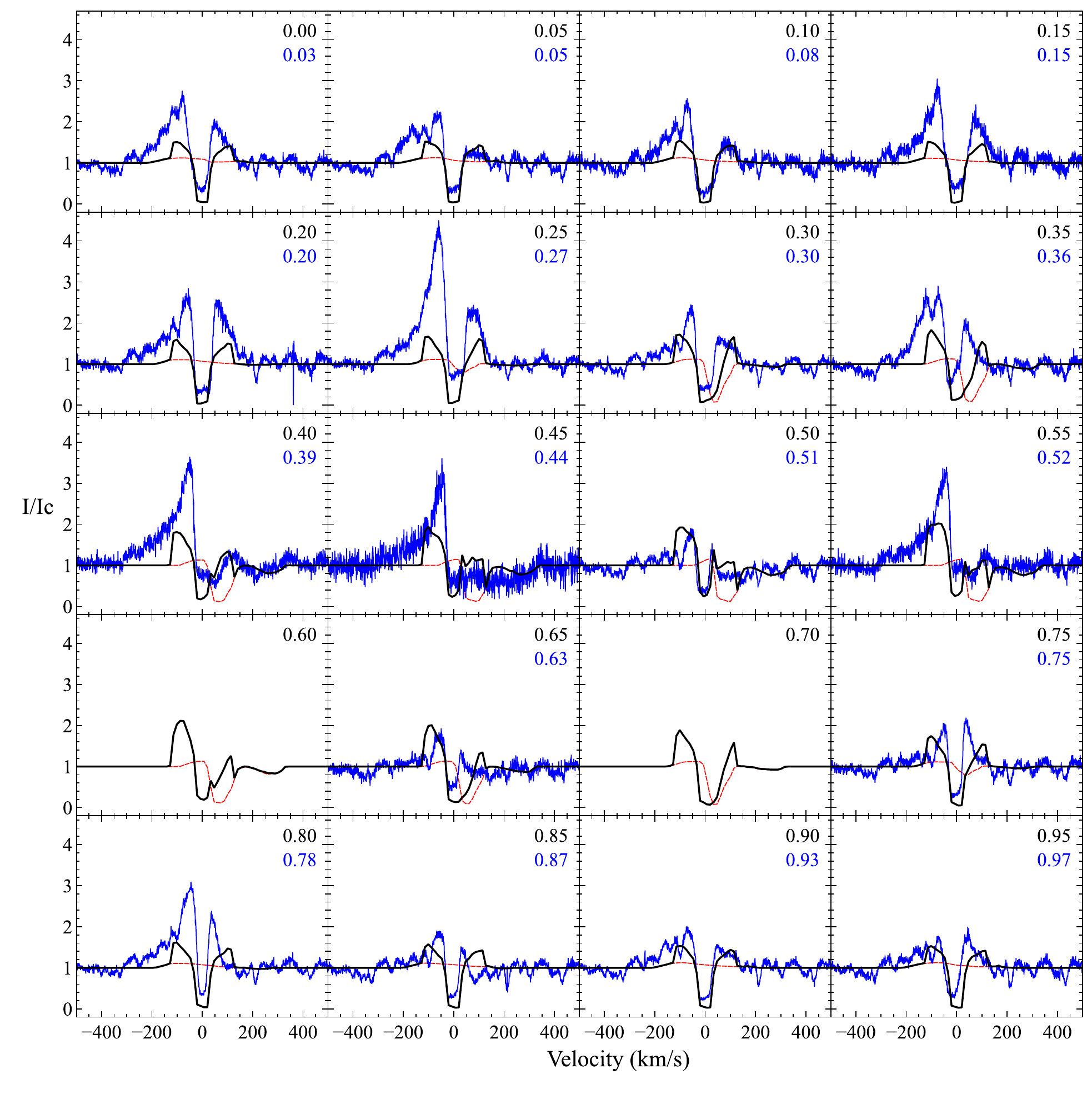}
\caption{Best fitting models for H{\gbeta} profiles during occultation, with $\dot{M}_{wind}/\dot{M}_{acc}=0.09$, $T_{mag}=$ 8500~K and $T_{wind}=$ 8000~K. The magnetosphere contribution is shown by the red dotted line, and the magnetosphere-plus-wind profile is shown by the black solid line.}
\label{linesfit_duringocc}
\end{figure*}

Out-of-occultation profiles are well-described by the canonical value of $\dot{M}_{wind}/\dot{M}_{acc}=0.1$, $T_{mag}=$ 8900~K and $T_{wind}=$ 8000~K. However, we have been unable to reduce the redward emission to match observations during occultation. The full grid of profiles for this model is shown in Fig.~\ref{linesfit_outofocc}. During occultation, the line profiles are generally best described by a model with a cooler magnetosphere, where $\dot{M}_{wind}/\dot{M}_{acc}=0.09$, $T_{mag}=$ 8500~K and $T_{wind}=$ 8000~K, shown in Fig.~\ref{linesfit_duringocc}. Individual profiles during occultation are also well-matched by $\dot{M}_{wind}/\dot{M}_{acc}=0.1$, $T_{mag}=$ 8500~K and $T_{wind}=$ 7900~K ($\varphi=0.5$, Fig.~\ref{R10M85W79All}) and $\dot{M}_{wind}/\dot{M}_{acc}=0.1$, $T_{mag}=$ 8400~K and $T_{wind}=$ 8000~K ($\varphi=$0.40 and 0.55, Fig.~\ref{R10M84W80All}).

\section{Discussion}
\label{sec:discussion}
The geometry of AA Tau is fairly well constrained by photometry, despite variations in the light curves between rotations. There is no degeneracy between the aspect ratio of the warp and the dipole offset. While light curves with a small aspect ratio and dipole offset are shallower than those with larger values, models with a smaller aspect ratio and larger offset are not matched by models with a larger aspect ratio and smaller offset because these parameters affect different parts of the light curve -- the out-of-occultation brightness appears to depend strongly on the the dipole offset, where larger offsets yield brighter maxima. This restricts the range in suitable dipole offset values, where aspect ratio is then varied to achieve a suitable depth during occultation. The inner magnetosphere radius has negligible effect on $V$-band photometry. $B$-band results have a much stronger dependence on this value, particularly for larger aspect ratios.

While we have used spectroscopic models to obtain a best fitting value for $\beta$, it is unlikely that spectroscopy places stronger constraints on the dipole offset angle than photometry in general due to the degeneracies present in the disc wind model. While we have presented two best fitting parameter sets in the spectroscopic study, there are numerous other acceptable solutions. In general, we have found that for the canonical value of $\dot{M}_{wind}/\dot{M}_{acc}\sim0.1$ with $T_{mag} = 8500$ -- 8900~K and $T_{wind} = 8000$~K are required.
A wind of some description is certainly necessary to produce the deep central absorption feature seen in observations. It is not clear whether the dominant source of outflow from CTTs is a disc wind, jet, stellar wind, or some combination of these, but our disc wind model matches the absorption feature fairly well.

The main issue with these results is the strength of the red emission peak during occultation. We have demonstrated that the red portion of the profile shows absorption features for sufficiently low mass-loss rates and temperatures, but we have been unable to produce profiles with both sufficient emission at phases out of occultation and sufficient damping on the red side during occultation to match observations. This is particularly noticeable in H{\galpha} during occultation, where both the red and blue peaks are significantly stronger than observations, despite providing a reasonable fit out of occultation (Figs B\ref{linesfit_outofoccHa} and B\ref{linesfit_duringoccHa}). There is also a discrepancy between the line broadening in models and observations for H{\gbeta}. While hotter magnetospheres produce more broadening (see Fig.~\ref{compareTmag}), these temperatures also produce significant redward emission, exacerbating the previous issue with our wind geometry.
As with H{\galpha}, our H{\ggamma} profiles fit observations fairly well out of occultation, but do not provide such a good description during occultation (Figs B\ref{linesfit_outofoccHg} and B\ref{linesfit_duringoccHg}), where in this case our model profiles do not produce such strong emission, even in the epochs where H{\gbeta} fits well (e.g. $\phi=0.5$).
This discrepancy may be rectified by using a more complex wind geometry; given the asymmetry of the accretion flow and the evidence for winds being powered by the accretion process, it may be natural to expect a similar asymmetry to be found in the wind parameters. The ``eggbeater" model of \cite[][their Fig. 15]{JohnsBasri95} may give a more accurate solution. This model describes a wind where rotational modulation occurs in $180^\circ$ phases since material is loaded more easily on to the wind flow when there is a large potential to overcome along the funnel flow, i.e. for $\phi=0$, with the reverse being true for $\phi=0.5$.

We have been able to simultaneously fit photometry and spectroscopy to AA Tau observations for H{\galpha}, H{\gbeta} and H{\ggamma} spectral lines and have recovered numerous parameters describing the entire system that are consistent with previous studies. Our study of the magnetosphere has found that photometric and spectroscopic observations are fitted well by a model of a dipole field with an offset of $\theta =$ 10 to 20$^\circ$, consistent with \citet{Donati10} who found that the magnetic field of AA Tau is best described by a dipole inclined at $\theta\simeq20^\circ$. \citet{Valenti04} derived an offset of $\theta=12^\circ$ from spectropolarimetric measurements, although they assumed a smaller inclination angle of $i=66^\circ$. We have also constrained the height of the disc warp, finding photometric data are best matched by an aspect ratio of $h_{max}/r_{o}=0.31$. This is in agreement with \citet{Alencar10}, who find $h_{max}/r_{o}\sim0.3$ is to be expected given that 28 per cent of their sample from $CoRoT$ observations show light curves similar to that of AA Tau, assuming a random distribution of system inclinations.
We also recover the canonical value of $\dot{M}_{wind}/\dot{M}_{acc} = 0.1$, with values of around 0.05 or lower producing insufficient emission at all rotational phases, and values greater than 0.15 producing excessive emission, particularly at $\varphi = 0.5$. However, we have found there is a degeneracy between the wind temperature and the acceleration parameter which cannot be constrained using the line profiles modelled here. One further area of interest is the He\,{\sc i} $\lambda$10830 line, found to be present for AA Tau in a survey by \citet{Edwards06}. He\,{\sc i} $\lambda$10830 acts more like a resonance line rather than the recombination lines we have considered here. With phase-resolved spectroscopy of the $\lambda$10830 line over AA Tau's 8 d period, we could potentially break this degeneracy.
Further constraints could be placed on the geometry of AA Tau with ground-based high-resolution imaging techniques such as interferometry. Direct imaging of AA Tau has already been achieved using the \emph{Hubble Space Telescope} clearly showing the disc, as well as showing evidence of jet outflows \citep{Cox13}. Photopolarimetry is another avenue of study for constraining the disc structure of AA Tau. Polarimetric variations have been observed by \citet{Bouvier99, Bouvier03} and \citet{Menard03}, who found that polarization is strongest when the system is faint, consistent with the presence of additional disc material enhancing the amount of polarization. Photopolarimetry of the warp modelled by \citet{OSullivan05} yields a warp position slightly interior to the corotation radius, with their warped disc model reproducing observed brightness and polarization variations well.

\section{Summary}
\label{sec:summary}
While initially variations in the light curves of CTTs were thought to be rare, such features have been shown to be present in at least 28 per cent of light curves \citep{Alencar10}. AA Tau was the first CTT observed to exhibit such photometric variations, and has been included in, and the subject of, many observational campaigns since \citep[e.g.][]{Gullbring98,Bouvier99,Bouvier07,Bouvier13,Grankin07,JohnsKrull07,Donati10}. Using the radiative transfer code {\sc torus} we have produced synthetic photometry and spectroscopy which were compared to photometric and spectroscopic observations from B99, B03 and B07, as well as photometry from \citetalias{ASAS}, to constrain geometric and physical parameters for AA Tau, simultaneously fitting photometry and spectroscopy. The geometry of the system varies over time as the magnetic field lines emanating from the star break and reconnect, so while our best fitting parameter sets have been found to best describe the system on average, different values will give superior fits at different times. Photometric models were used to constrain the mass accretion rate, the maximum height of the inner disc warp, the dipole offset, and the inner radius of the magnetosphere. Spectroscopic models were used to further constrain the dipole offset, and to investigate the mass-loss rate from a disc wind, the temperature of the wind, and the temperature of the magnetosphere. We found that models with a mass accretion rate of $\sim 5 \times 10^{-9}$ M$_{\odot}$ yr$^{-1}$ yielded an accretion luminosity of up to $6.1 \times 10^{-2}$ L$_{\odot}$, consistent with the value of $6.5 \times 10^{-2}$ L$_{\odot}$ calculated by \citetalias{Bouvier99}. This accretion rate is consistent with the range of 2--7 $\times 10^{-9}$ M$_{\odot}$ yr$^{-1}$ calculated from H{\galpha} and H{\gbeta} line fluxes in \citet{Bouvier13}. We collated all publicly available $B$- and $V$-band photometry and found average light curves are best described by $h_{max}/r_{o}\sim0.31$ and $\theta=10^\circ$--$20^\circ$, consistent with B99 who also derived $h_{max}/r_{o}=0.3$ \cite[see also][]{Terquem00}, and \citet{Alencar10} who also found $h_{max}/r_{o}\sim0.3$ for AA Tau-like light curves. The range in dipole offsets is consistent with \citet{Donati10} who found $\theta=20^\circ$, although our spectroscopic models favour $\theta=10^\circ$. A disc wind is required to recover the line shapes observed. Although degeneracies between density and temperature in the disc wind prevent the determination of absolute values, we have found that the canonical value for mass-loss rate to mass accretion rate of 0.1 is recovered. Disc wind and magnetosphere temperatures of $T_{wind}\sim$ 7900--8000 K and $T_{mag}\sim$ 8400--8500 K have been established.

A future study of the He\,{\sc i} $\lambda$10830 line for AA Tau would be beneficial to further constrain the outflow mechanism. Developing a more complex, azimuthally asymmetric disc wind calculation in {\sc torus} may also account for the variations between the best fitting wind parameters at different phases encountered here.
However, we cannot progress much further than this by fitting spectra alone.
Extra leverage may soon be achievable with spectrally resolved interferometry, which has already been used successfully to infer information about Herbig Ae/Be stars \citep*[e.g.][]{Kraus07,Kraus08}. Current instrumentation is unable to achieve the sensitivity in both signal-to-noise and resolution to probe the small inner region of T Tauri stars, but it remains an exciting prospect for the next generation of interferometers.
The advent of high-resolution interferometry will allow us to infer spatial information about CTT systems, providing much stronger constraints on a system's geometry.

\section*{Acknowledgements}
The calculations for this paper were performed using the University of Exeter Supercomputer. We thank the referee for their comments. CFE acknowledges funding from a College studentship awarded by The Univeristy of Exeter, and TJH acknowledges funding from the STFC grant ST/J001627/1. Finally, we acknowledge the $ASAS$ project for additional AA Tau photometric data (available online at {\tt http://www.astrouw.edu.pl/asas/}).

\appendix
\section[]{Comparisons between observations and synthetic photometry}
\label{appendixA}
Here we present some of our photometric models with averaged observational photometry, for $B$-band (top) and $V$-band (bottom). Different magnetosphere sizes are presented, with $r_i = 5.2$ (dashed), $r_i = 6.4$ (dotted), and $r_i = 7.6$ (dash-dot). Dipole offset and aspect ratio have also been varied, with values for each of these given in the captions. A mass accretion rate of $\dot{M}_{acc} = 5\times10^{-9}$~M$_\odot$~yr$^{-1}$ is used. All other parameters are as presented in Table \ref{tab:params}, in the main text.

\floatplacement{figure}{H}
\begin{figure}
\includegraphics[width=85mm]{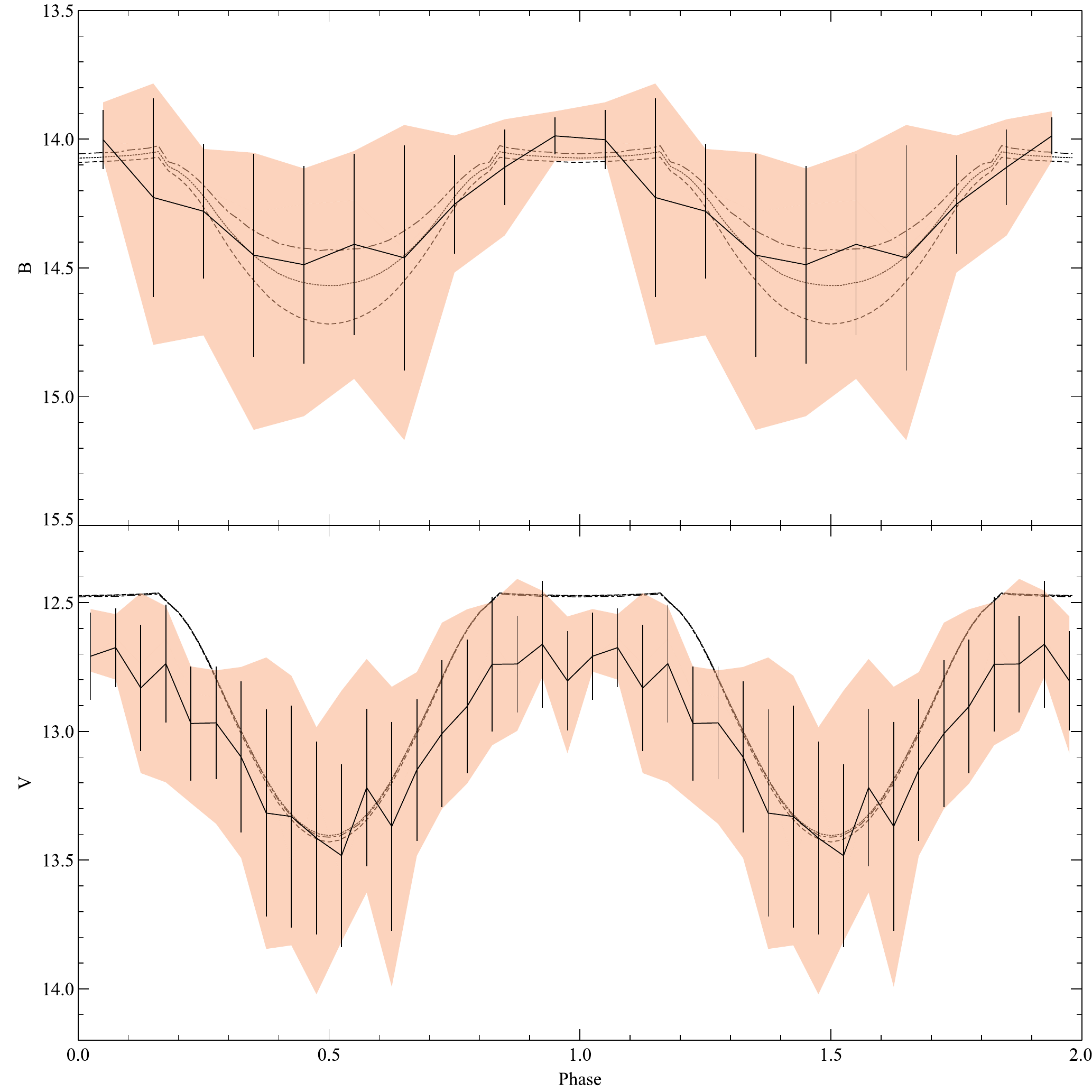}
\caption{Comparisons between observations and models where $\theta = 10^\circ$ and $h_\mathrm{max}/r_{o} = 0.31$.}
\label{o10h31}
\end{figure}

\begin{figure*}
\centering
\begin{minipage}[b]{0.48\linewidth}
	\includegraphics[width=85mm]{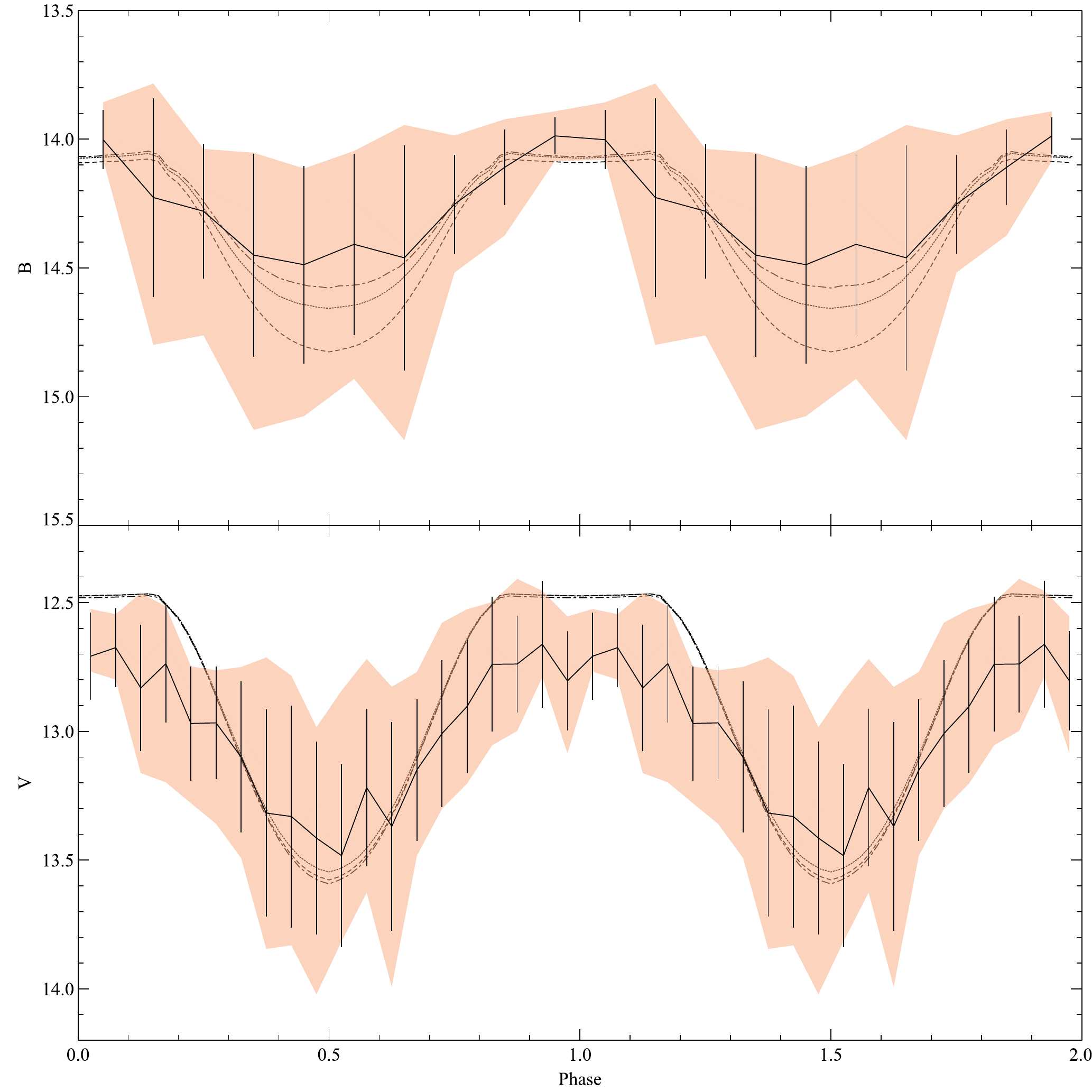}
	\caption{Comparisons between observations and models where $\theta = 10^\circ$ and $h_\mathrm{max}/r_{o} = 0.32$.}
	\label{o10h32}
\end{minipage}
\quad
\begin{minipage}[b]{0.46\linewidth}
	\includegraphics[width=85mm]{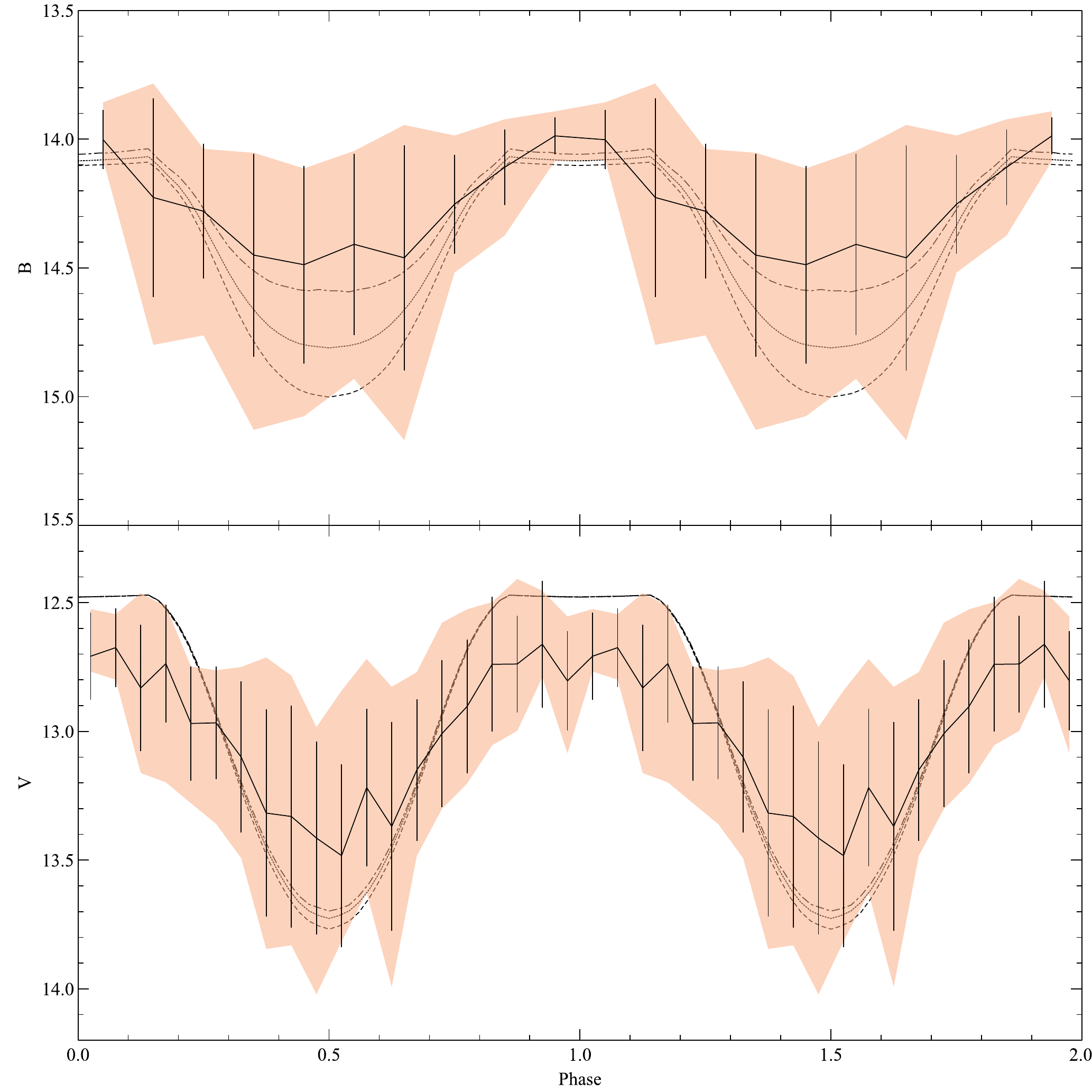}
	\caption{Comparisons between observations and models where $\theta = 10^\circ$ and $h_\mathrm{max}/r_{o} = 0.33$.}
	\label{o10h33}
\end{minipage}
\end{figure*}

\begin{figure*}
\centering
\begin{minipage}[b]{0.48\linewidth}
	\includegraphics[width=85mm]{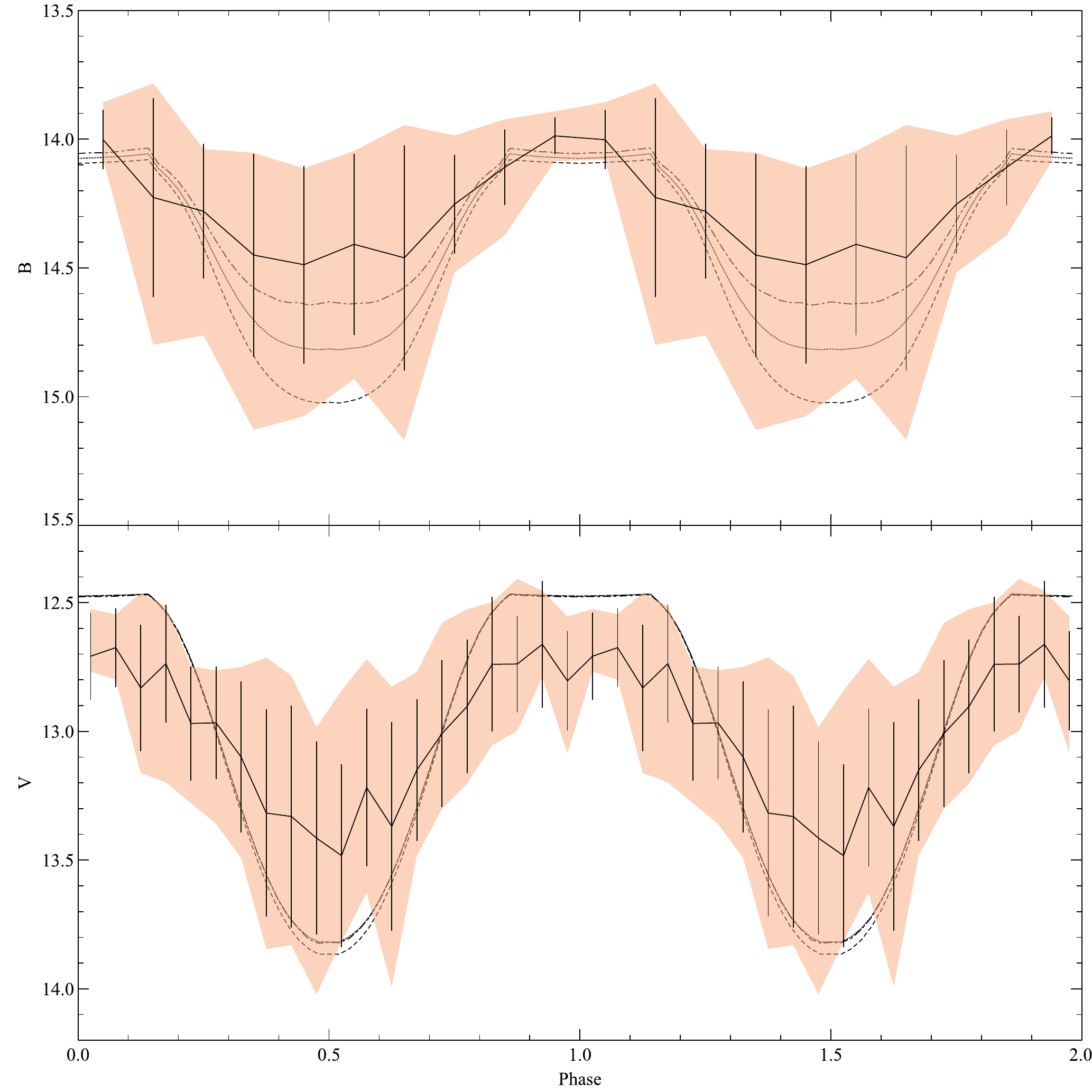}
	\caption{Comparisons between observations and models where $\theta = 10^\circ$ and $h_\mathrm{max}/r_{o} = 0.34$.}
	\label{o10h34}
\end{minipage}
\quad
\begin{minipage}[b]{0.46\linewidth}
	\includegraphics[width=85mm]{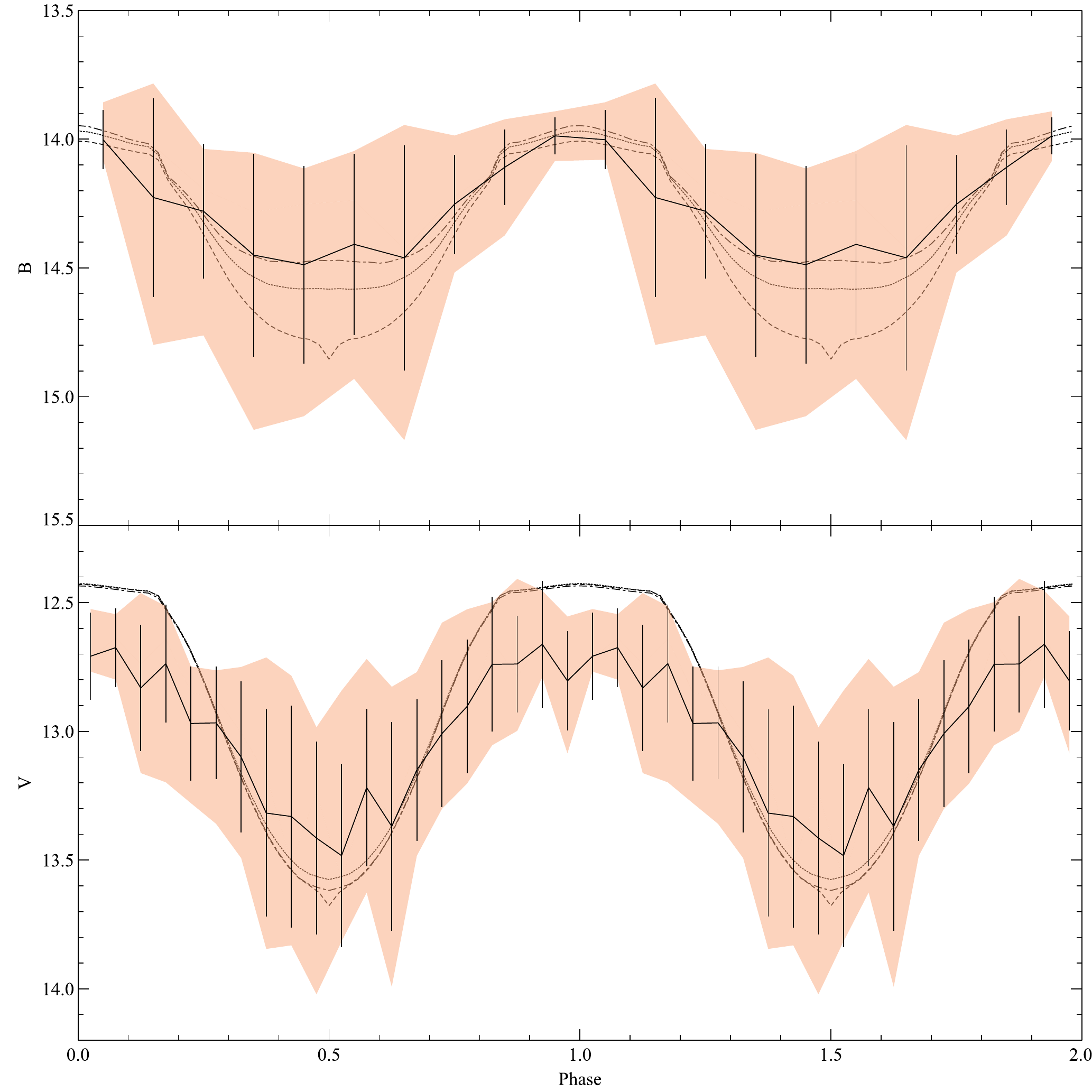}
	\caption{Comparisons between observations and models where $\theta = 20^\circ$ and $h_\mathrm{max}/r_{o} = 0.33$.}
	\label{o20h33}
\end{minipage}
\end{figure*}

\begin{figure*}
\centering
\begin{minipage}[b]{0.48\linewidth}
	\includegraphics[width=85mm]{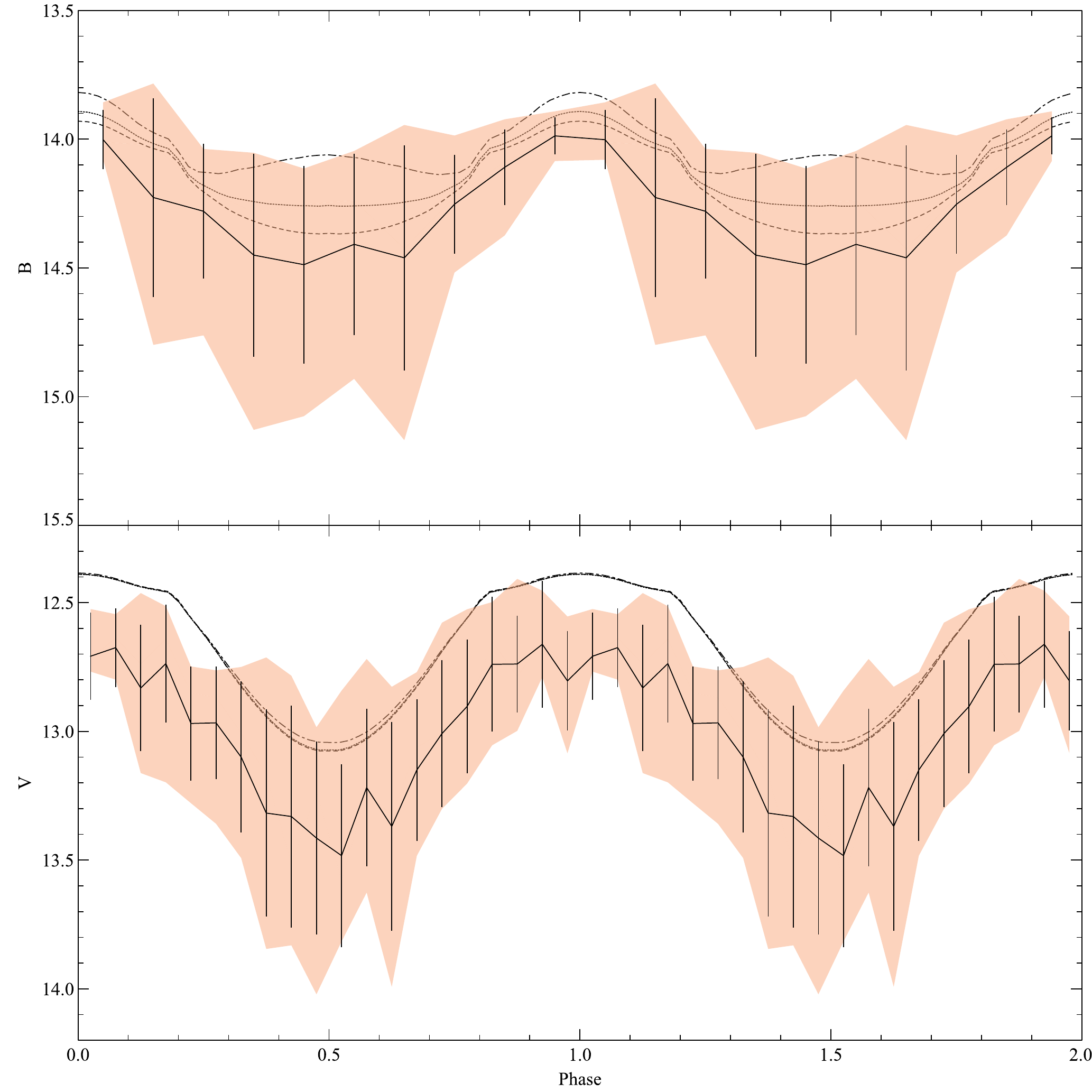}
	\caption{Comparisons between observations and models where $\theta = 30^\circ$ and $h_\mathrm{max}/r_{o} = 0.29$.}
	\label{o30h29}
\end{minipage}
\quad
\begin{minipage}[b]{0.46\linewidth}
	\includegraphics[width=85mm]{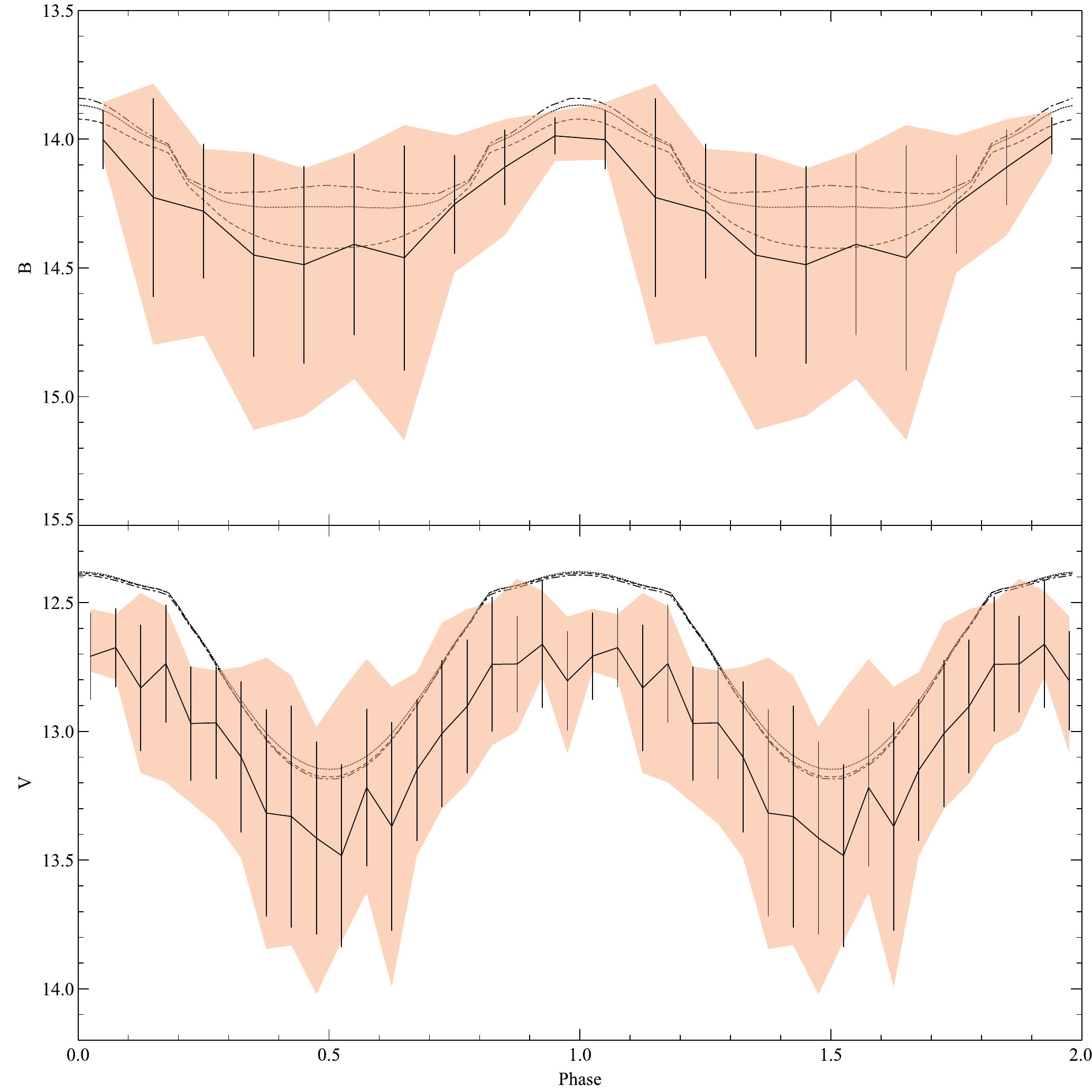}
	\caption{Comparisons between observations and models where $\theta = 30^\circ$ and $h_\mathrm{max}/r_{o} = 0.30$.}
	\label{o30h30}
\end{minipage}
\end{figure*}

\begin{figure*}
\centering
\begin{minipage}[b]{0.48\linewidth}
	\includegraphics[width=85mm]{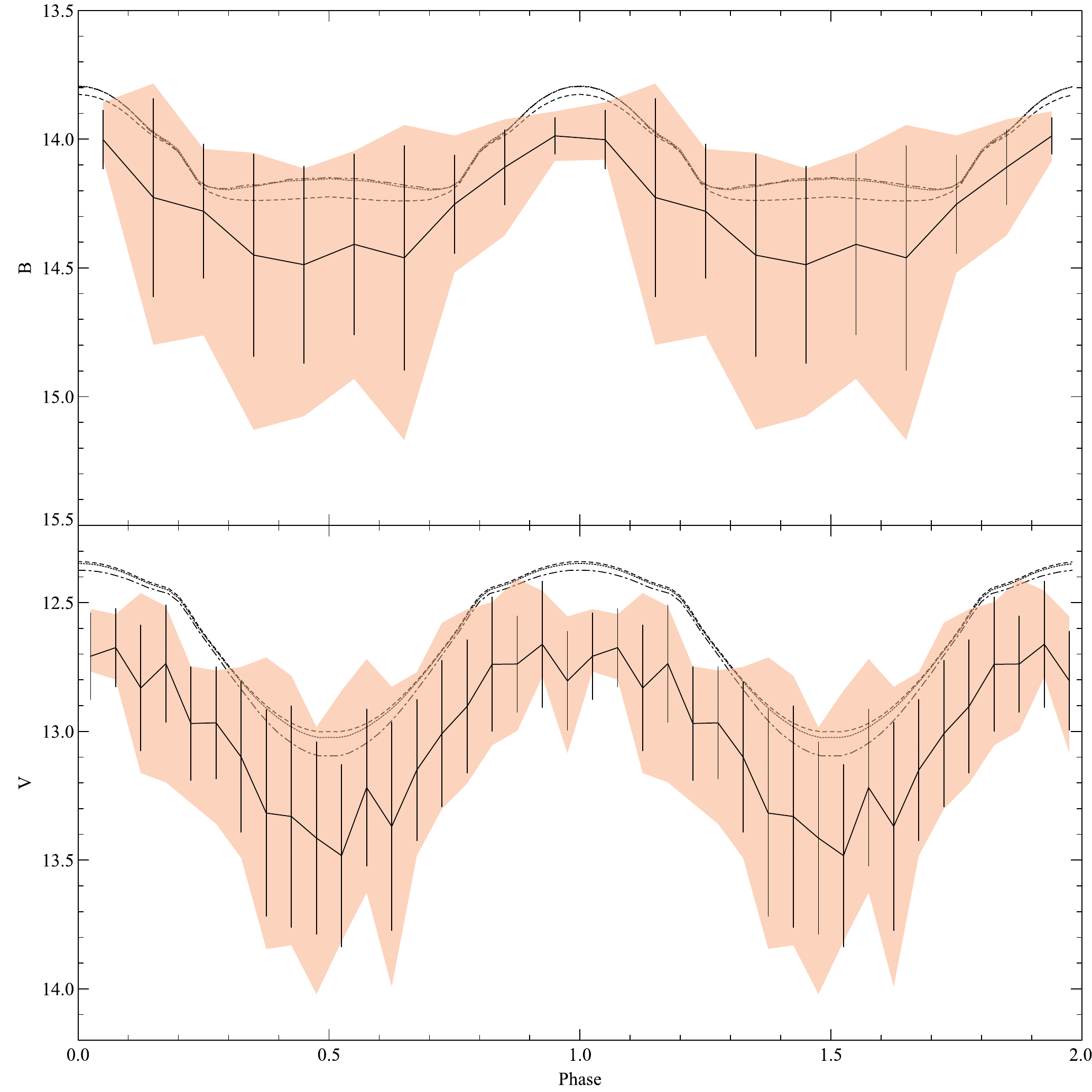}
	\caption{Comparisons between observations and models where $\theta = 40^\circ$ and $h_\mathrm{max}/r_{o} = 0.29$.}
	\label{o40h29}
\end{minipage}
\quad
\begin{minipage}[b]{0.46\linewidth}
	\includegraphics[width=85mm]{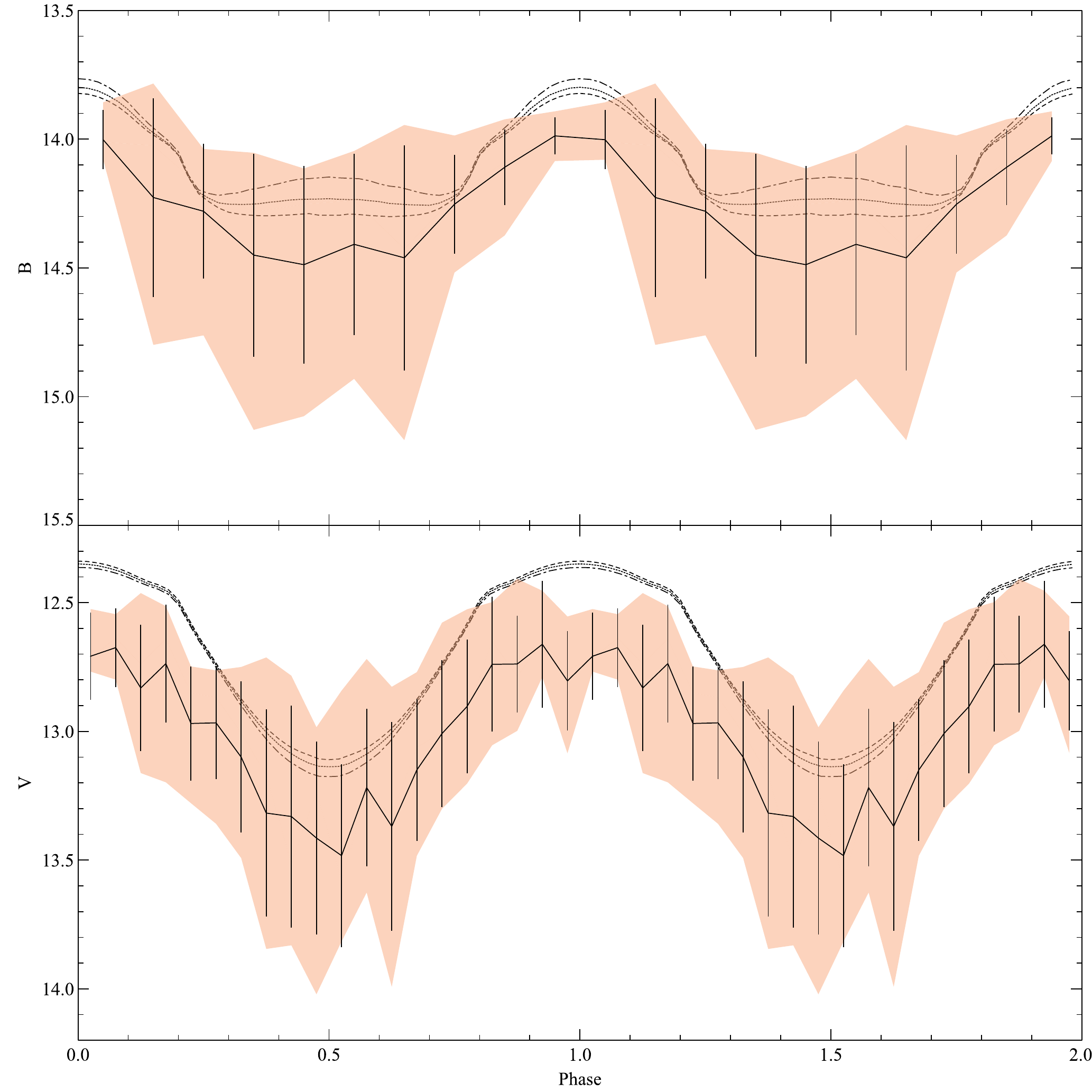}
	\caption{Comparisons between observations and models where $\theta = 40^\circ$ and $h_\mathrm{max}/r_{o} = 0.30$.}
	\label{o40h30}
\end{minipage}
\end{figure*}

\section[]{Comparisons between observations and synthetic spectroscopy}
\label{appendixB}
Here we present spectroscopic models with observed line profiles for H{\galpha}, H{\gbeta} and H{\ggamma}. A selection of models are presented, as referenced in the main body of the text. The ratio of mass-loss rate to mass-accretion rate, the wind temperature, and the magnetosphere temperature have been varied, with values for these given in the figure captions. All other parameters are as presented in Table \ref{tab:windparams}, in the main text.

\begin{figure*}
\centering
\mbox{\hspace{-5mm} \subfigure[H{\galpha}]{\includegraphics[width=96mm]{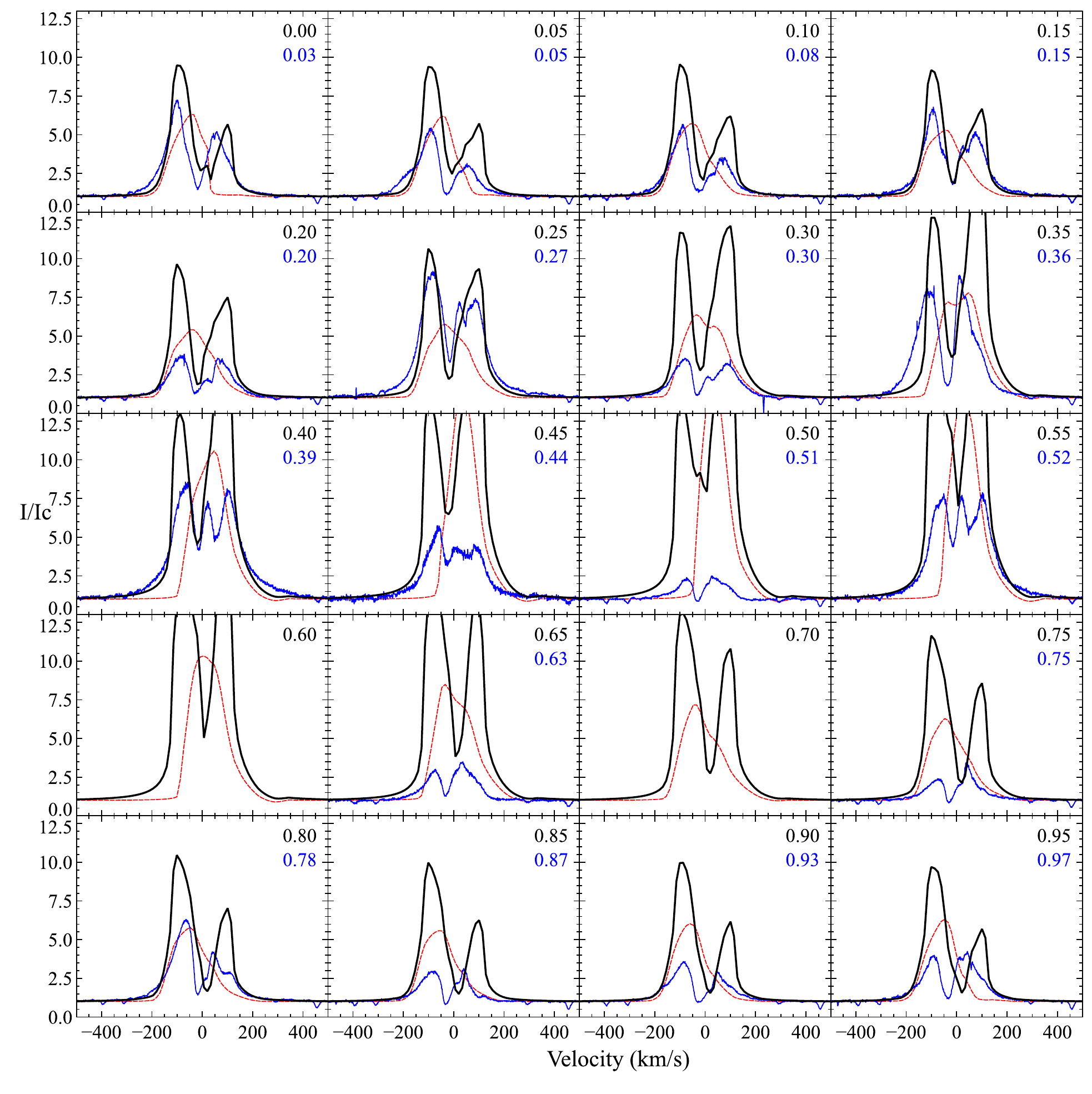}\label{linesfit_outofoccHa}}
\quad
\subfigure[H{\ggamma}]{\hspace{-5mm} \includegraphics[width=96mm]{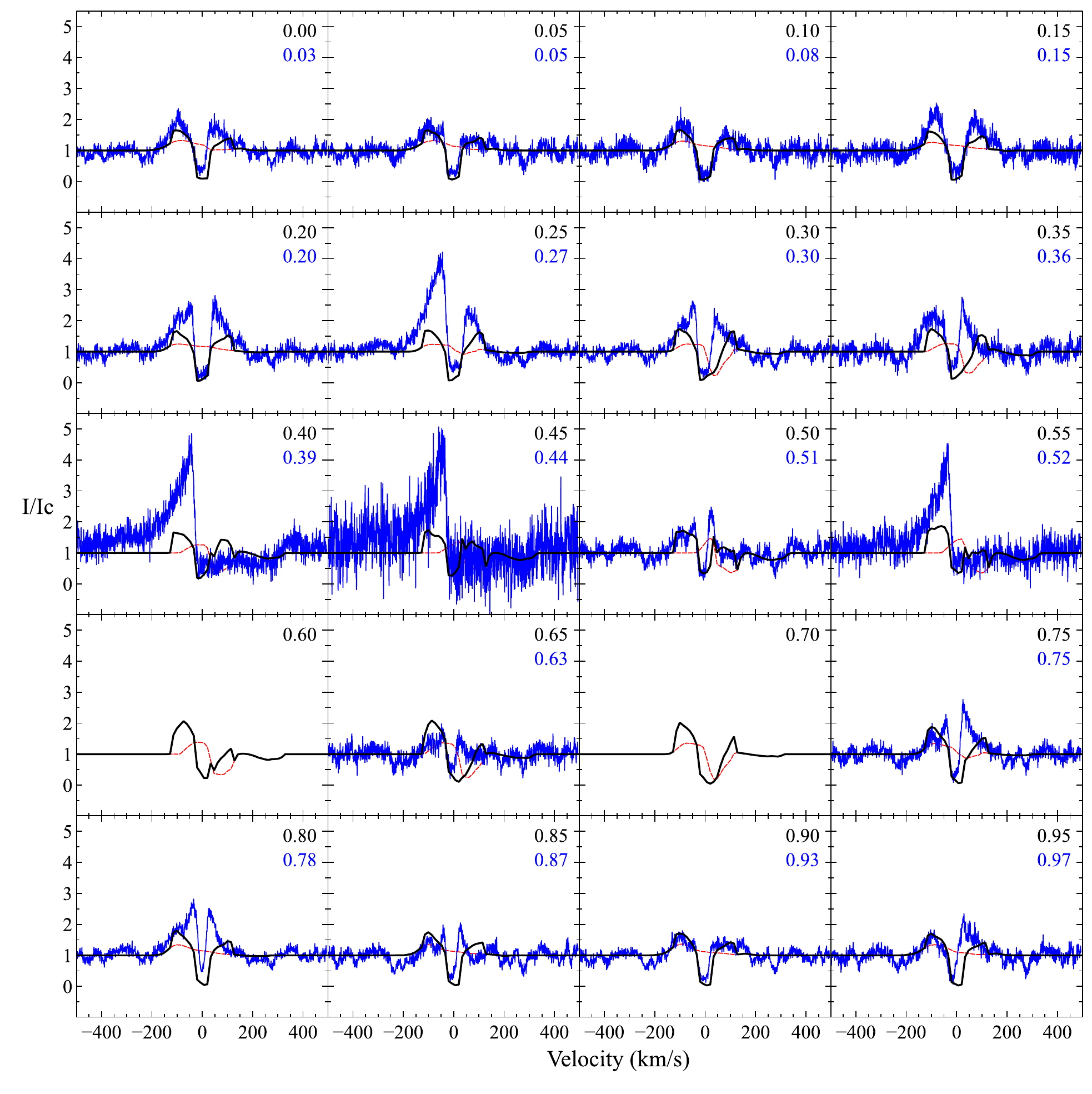}\label{linesfit_outofoccHg}}}
\caption{Comparison of models with observed line profiles for $\dot{M}_{wind}/\dot{M}_{acc}=0.1$, $T_{mag}=$ 8900~K and $T_{wind}=$ 8000~K. The magnetosphere contribution is shown by the red dotted line.}
\label{linesfit_outofoccHaHg}
\end{figure*}

\begin{figure*}
\centering
\mbox{\hspace{-5mm} \subfigure[H{\galpha}]{\includegraphics[width=96mm]{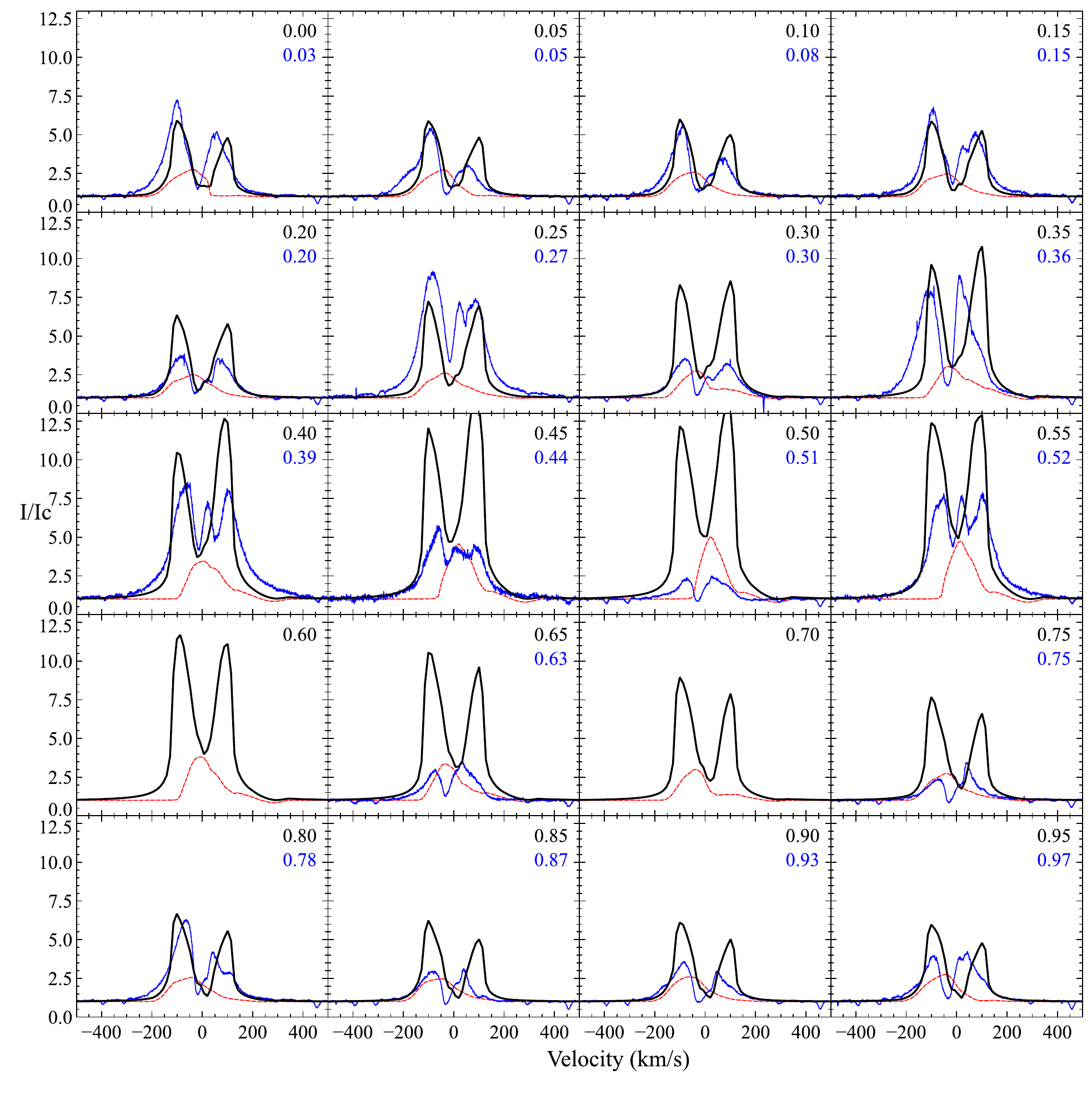}\label{linesfit_duringoccHa}}
\quad
\subfigure[H{\ggamma}]{\hspace{-5mm} \includegraphics[width=96mm]{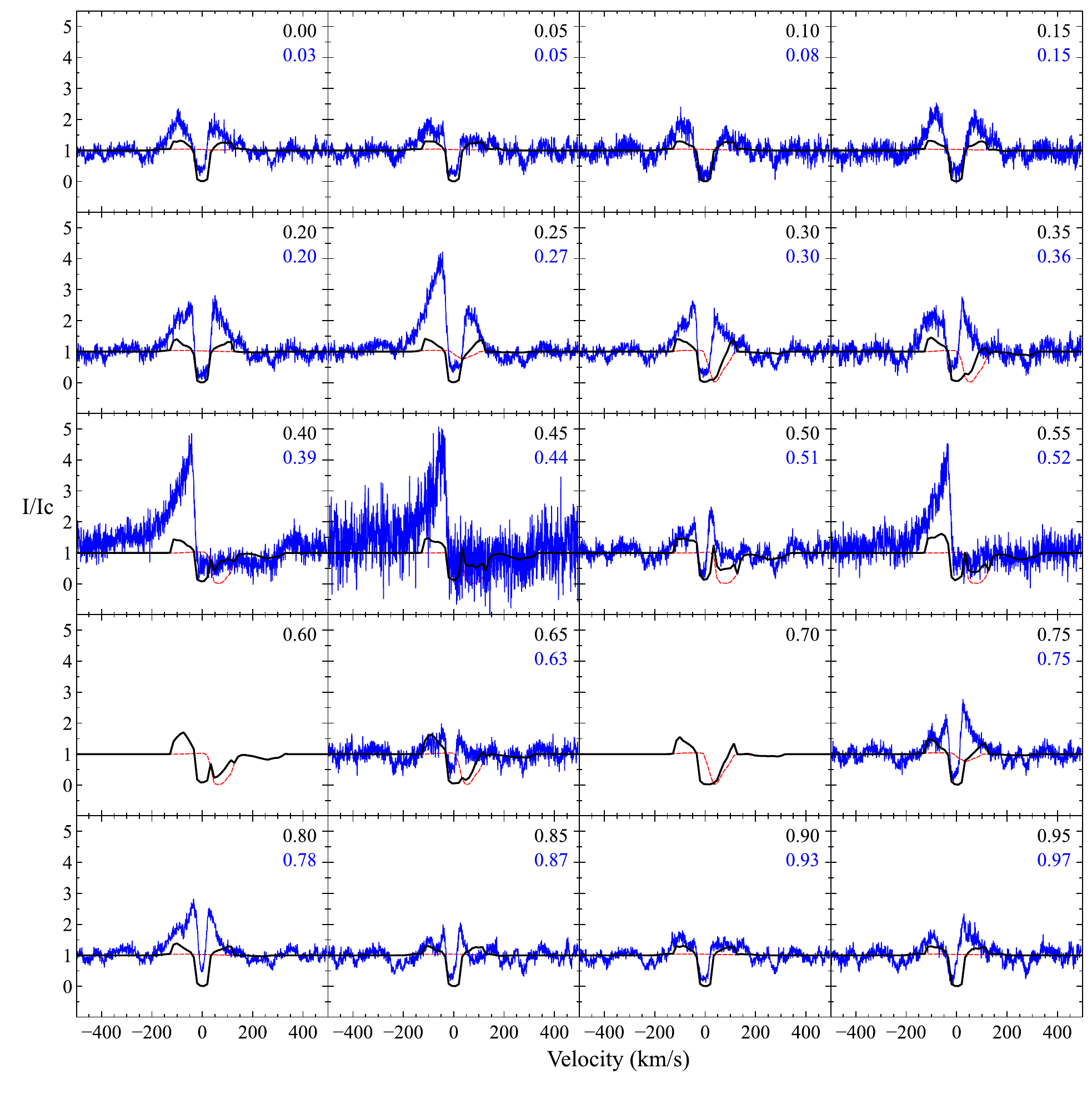}\label{linesfit_duringoccHg}}}
\caption{Comparison of models with observed line profiles for $\dot{M}_{wind}/\dot{M}_{acc}=0.09$, $T_{mag}=$ 8500~K and $T_{wind}=$ 8000~K for H{\galpha} and H{\ggamma}. The magnetosphere contribution is shown by the red dotted line.}
\label{linesfit_duringoccHaHg}
\end{figure*}

\begin{figure*}
\centering
\includegraphics[width=165mm]{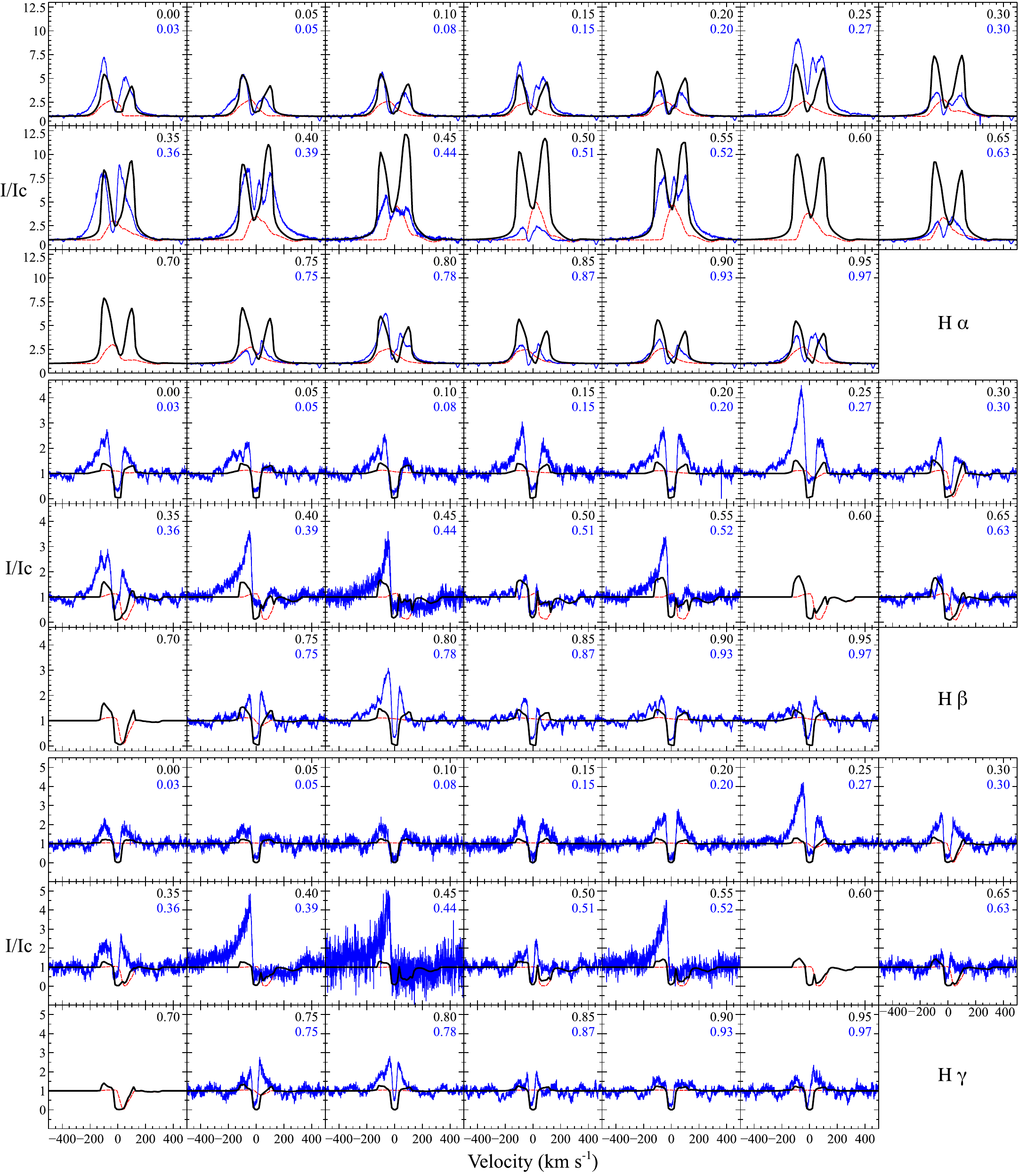}
\caption{Observed H{\galpha} , H{\gbeta} and H{\ggamma} profiles compared with models where $\dot{M}_{wind}/\dot{M}_{acc}=0.10$, $T_{mag}=$ 8500~K and $T_{wind}=$~7900~K. The magnetosphere contribution is shown by the red dotted line.}
\label{R10M85W79All}
\end{figure*}

\begin{figure*}
\centering
\includegraphics[width=165mm]{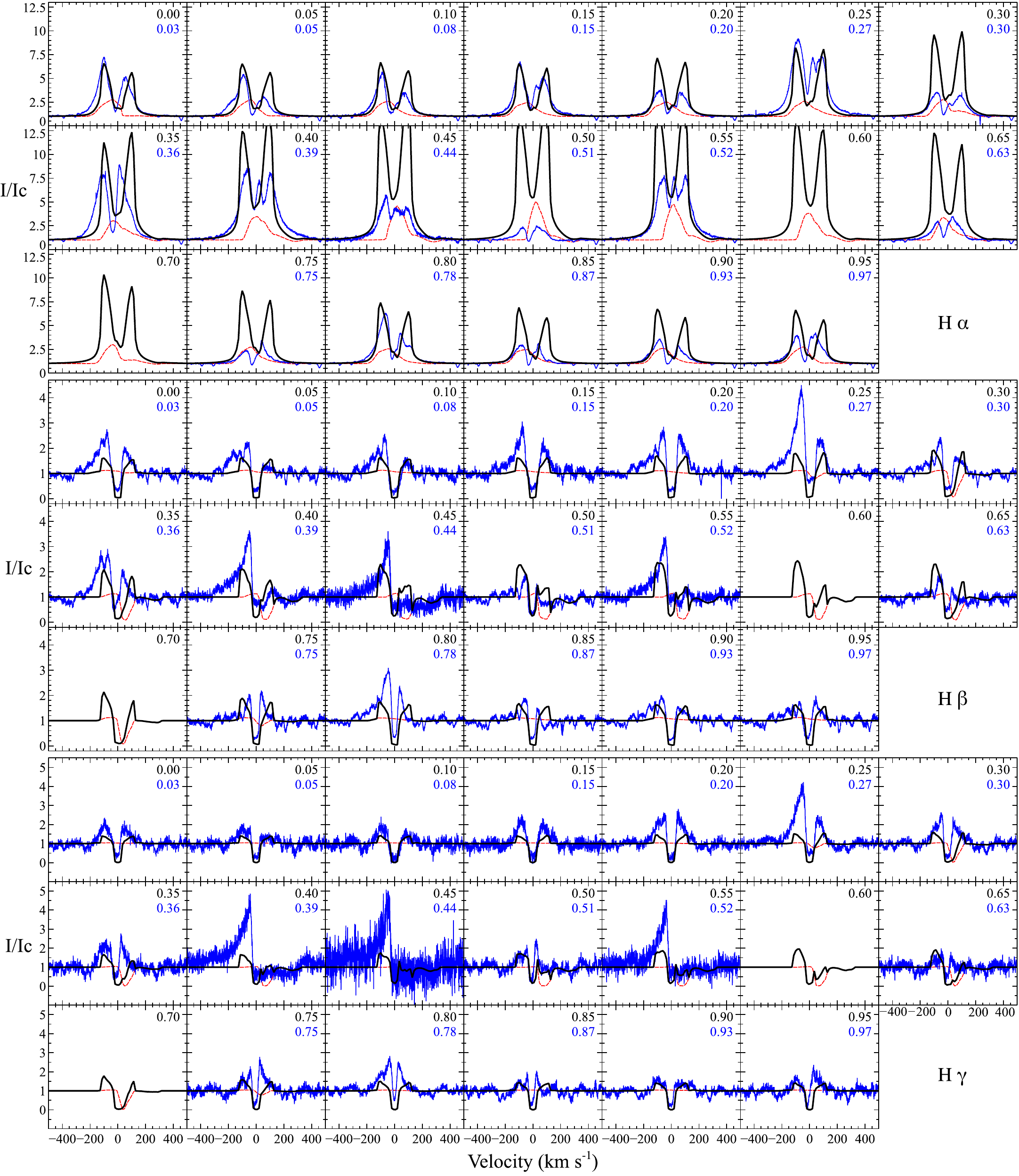}
\caption{Observed H{\galpha} , H{\gbeta} and H{\ggamma} profiles compared with models where $\dot{M}_{wind}/\dot{M}_{acc}=0.10$, $T_{mag}=$ 8400~K and $T_{wind}=$~8000~K. The magnetosphere contribution is shown by the red dotted line.}
\label{R10M84W80All}
\end{figure*}

\bsp
\label{lastpage}

\end{document}